\documentclass[12pt]{article}

\input{preamble}
\usepackage[margin=1in]{geometry}
\usepackage{natbib}
\usepackage{hyperref}
\hypersetup{colorlinks=true, linkcolor=blue, urlcolor=blue, citecolor=blue}
\usepackage{setspace}
\usepackage{amsmath}
\newtheorem{theorem}{Theorem}
\newtheorem{lemma}{Lemma}
\newtheorem{proof}{Proof}

\usepackage{enumerate}
\newtheorem{example}{Example}
\newtheorem{remark}{Remark}
\renewcommand{\ss}{s}

\begin{document}

\title{Asymptotic Validity and Finite-Sample Properties of Approximate Randomization Tests}

\author{Panos Toulis \\Booth School of Business, University of Chicago
\thanks{{\em Email:} panos.toulis@chicagobooth.edu; {\em Addr.:} 5807 S Woodlawn Ave, Chicago, IL 60637, USA}}


\maketitle

\begin{abstract}
Randomization tests rely on simple data transformations and possess an appealing robustness property. In addition to being finite-sample valid if the data distribution is invariant under the transformation, these tests can be asymptotically valid under a suitable studentization of the test statistic,  even if the invariance does not hold. However, practical implementation often encounters noisy data, resulting in {\em approximate} randomization tests that may not be as robust. In this paper, our key theoretical contribution is a non-asymptotic bound on the discrepancy between the size of an approximate randomization test and the size of the original randomization test using noiseless data. This allows us to derive novel conditions for the validity of approximate randomization tests under data invariances, while being able to leverage existing results based on studentization if the invariance does not hold. We illustrate our theory through several examples, including tests of significance in linear regression. Our theory can explain certain aspects of how randomization tests perform in small samples, addressing limitations of prior theoretical results. 
\end{abstract}


\onehalfspacing
\section{Introduction}
\label{sec:introduction}
In many statistical applications, we need to test null hypotheses on data  $\eps = (\eps_1, \ldots, \eps_n) \in \Real^n$  that imply an invariance 
property of the form:
\begin{equation}\label{eq:invariant_hypothesis}
\eps \myeq{d} \gg\eps,~\text{for all}~\gg\in \Gall, n > 0,
\end{equation}
where $\Gall$ is a finite group of $\mathbb{R}^n\to\mathbb{R}^n$ transformations.
Such a null hypothesis can be readily tested using the randomization method~\citep{fisher1935design,lehmann2006testing}. 
The idea is to choose a test statistic $T_n$ such that, under the null hypothesis, $T_n = t_n(\eps)$ for a known measurable function $t_n:\Real^n\to\Real$. Then, the randomization test compares the observed value of the test statistic to the values obtained by transforming the data according to the invariance:
\begin{equation}\label{eq:phi_star}
\phi_n^*= \Indb{T_n > c_{n, \alpha} (\eps) }.
\end{equation}

Here, $c_{n, \alpha}(u) =  \inf\{ z \in\Real: \sum_{\gg\in\Gall} \Ind\{t_n(\gg u) \le z\} \ge |\Gall|(1-\alpha)\}$ is the critical value function, and $\alpha$ is the level of the test. Under the {\em invariant hypothesis} described by Equation~\eqref{eq:invariant_hypothesis}, $\phi_n^*$ is valid in finite samples for any choice of test statistic~\citep[Theorem~15.2.3]{lehmann2006testing}, requiring no further assumptions on the distribution of $\eps$ beyond the invariance. This has made randomization tests particularly popular in experimental design and causal inference, where robustness is key~\citep[Ch.~6]{cox1979theoretical};~\citep{rubin1980randomization, eaton1989group, ernst2004permutation, higgins2004introduction, hinkelmann2007design, edgington2007randomization, gerber2012field, good2013permutation};~\citep[Ch.~2]{rosenbaum2002observational};~\citep[Ch.~5]{imbens2015causal}.

Beginning from the seminal work of~\citet{janssen1997studentized}, many researchers have pointed out \citep{chung2013exact, diciccio2017robust, canay2017randomization, wu2021randomization, zhao2021covariate} that the randomization test can also be valid in settings that do not rely on an invariance hypothesis, but instead rely on a {\em limit hypothesis} of the form:
\begin{equation}\label{eq:limit_hypothesis}
t_n(\eps) := s_n(\eps) / \sigma_n(\eps) \limeq{d} \mathcal{L}.
\end{equation}
Here, $\sigma_n : \mathbb{R}^n \to \mathbb{R}^+$ is a 
standard error estimate of statistic $s_n(\eps)$, and $\mathcal{L}$ denotes a parameter-free probability law.
A classical theoretical result, due to \citet{hoeffding1952}, implies that if 
\begin{equation}\label{eq:hoeff}
(t_n(G\eps), t_n(G'\eps)) \limeq{d} (T, T')~,
\end{equation}
where $G, G'\sim\Unif(\Gall)$ and $T,T' \sim \mathcal{L}$, all mutually independent, then the randomization test $\phi_n^*$ can be asymptotically valid under the limit hypothesis described in Equation~\eqref{eq:limit_hypothesis} even if $\eps$ are not invariant under $\Gall$. Equation~\eqref{eq:hoeff} is known as `Hoeffding's condition'~\citep[Sec. 15.2.2]{lehmann2006testing}, and remains a key theoretical tool with which to study the asymptotic behavior of randomization tests. In essence, this condition requires that the randomization distribution of a test statistic and the statistic's sampling distribution ---i.e., random variables $t_n(G\eps)$ and $t_n(\eps)$, respectively--- are asymptotically identical.  Randomization tests therefore possess an appealing robustness property: a randomization test is always finite-sample valid under the invariant hypothesis~\eqref{eq:invariant_hypothesis}, and the same test is also asymptotically valid under the limit hypothesis~\eqref{eq:limit_hypothesis}.

However, in many practical settings, the random variables $\eps$ cannot be fully observed; e.g., in a regression model, $\eps$ may represent unobserved noise.  A standard approach to address this problem is to use  proxy variables, $\heps$ (e.g., regression residuals), leading to the definition of an {\em approximate randomization test},
\begin{equation}\label{eq:phi}
\phi_n = \Indb{T_n > c_{n,\alpha} (\hspace{1px}\heps\hspace{1px})}.
\end{equation}

While intuitive, the approximate test may no longer have the robustness properties of the original randomization test using the true variables. To study this problem,  a standard approach is to establish Hoeffding's condition on the asymptotic limit of $t_n(G\heps)$, i.e., the randomization distribution under the proxy variables.  However, this approach has opened up two notable theoretical gaps. First, while standard tools can establish the asymptotic validity of an approximate randomization test under the limit hypothesis, they do not provide insights into its performance under the invariant hypothesis. Furthermore, as current tools are based on asymptotic theory, they fail to provide insights into the finite-sample performance of randomization tests in general. This paper aims to address these existing theoretical gaps, and demonstrate the utility of its new theory through practical examples including linear regression.

\subsection{Overview of main results}\label{sec:overview}
The main result in this paper is a finite-sample bound  
between the sizes of the approximate randomization test and the original test.
That is, for some fixed constant $K>0$, we prove that
\begin{align}\label{eq:Main_Result}
|\mE(\phi_n) - \mE(\phi_n^*)| \le K \frac{\mE [ (t_n(G\heps) - t_n(G\eps) )^2 ]}{\mE[ (t_n(G' \eps) - t_n(G''\eps))^2]} + o(1),~~G, G', G'' \sim \Unif(\Gall)~\text{i.i.d.}
\end{align}
 The key implication of this result is that the approximate randomization test `inherits' the asymptotic properties of the original randomization test as long as 
\begin{align}\label{eq:C1}
\frac{\mE [ (t_n(G\heps) - t_n(G\eps) )^2 ]}{\mE[ (t_n(G' \eps) - t_n(G''\eps))^2]} \to 0.~\tag{C1}
\end{align}
%

This result offers a new theoretical tool with which to study approximate randomization tests in addition to what is currently possible through Hoeffding's condition. In particular, Condition~\eqref{eq:C1} does not require that the test statistic has a regular asymptotic limit, or that $t_n(G'\eps)$ and $t_n(G''\eps)$ are independent. 
The condition only stipulates that the variation in the error from the approximate randomization test using the proxy variables~(numerator) is dominated by the variation in the spacings of the statistic values in the original randomization test using the true variables~(denominator).
By way of illustration, in Section~\ref{sec:hoeff_compare} we will discuss an example where Hoeffding's condition fails but the approximate test remains valid in line with our theory.

Condition~\eqref{eq:C1} has another important implication: it shows that approximate randomization tests possess a robustness property that is similar, albeit weaker, to the robustness of true randomization tests. Specifically, the condition by itself guarantees the asymptotic validity of an approximate randomization test if the invariant hypothesis~\eqref{eq:invariant_hypothesis} is true.
Moreover, under the limit hypothesis, Condition~\eqref{eq:C1} guarantees the asymptotic validity of the approximate test provided that the studentized test statistic based on the true variables satisfies Hoeffding's condition. 

 Our analysis is further extended to study the finite-sample performance of approximate randomization tests across both hypothesis regimes.
Indeed,  Theorem~\ref{thm:main} of this paper shows that the rate of convergence of Condition~\eqref{eq:C1} determines a finite-sample bound between the Type I error rates of $\phi_n$ and $\phi_n^*$. This bound also depends on the smoothness constant of the cdf of the `spacings variable' in the denominator of~\eqref{eq:C1}. Intuitively, the approximate randomization test performs as well as the true test unless the multiplicity of values in the randomization distribution is too erratic. For example, if the test statistic is asymptotically normal, the bound is of order $O(n^{-1/3})$, suggesting a robustness-efficiency trade-off that we discuss throughout the paper. These results are especially valuable under the invariant hypothesis~\eqref{eq:invariant_hypothesis} as prior randomization literature has not studied the performance of approximate randomization tests under the invariant regime.

In Section~\ref{sec:linear}, we show how Condition~\eqref{eq:C1} can be simplified when testing the significance of coefficients in linear regression. In this context, we study randomization tests on regression residuals, which we will refer to as residual randomization tests. These procedures bear strong similarities to permutation tests of significance~\citep{janssen1997studentized, diciccio2017robust} and several bootstrap variants~\citep{freedman1983nonstochastic, wu1986jackknife, davidson2008wild}. 
Our theory provides simple conditions for the asymptotic validity of residual randomization tests under the invariant hypothesis, while still being able to leverage existing theoretical results under the limit hypothesis. As a useful byproduct, our analysis clarifies the trade-offs between data invariance assumptions of the form described by Equation~\eqref{eq:invariant_hypothesis} and i.i.d. assumptions underlying  classical bootstrap theory.

To give a concrete example, suppose that $\Gall$ is the symmetry group of permutations of $n$ elements. Then, Condition~\eqref{eq:C1} reduces to the remarkably simple $p/n = o(1)$ per Theorem~\ref{thm:perm}, for the corresponding residual permutation test of significance. This result has a twofold interpretation.  Firstly, if the regression errors are exchangeable, then the residual test is asymptotically valid if $p/n=o(1)$. Note that this places no additional restrictions on the distribution or asymptotics of $(X, \eps)$ beyond exchangeability, and allows for settings with high leverage and even high dimensions with $p<n$ but $p\to\infty$.  This result highlights the strength of data invariance assumptions of the form~\eqref{eq:invariant_hypothesis}. Secondly, if the invariant hypothesis does not hold, condition $p/n=o(1)$ implies the asymptotic equivalence between the residual permutation test and the permutation test using the true errors.  Hence, even in the absence of exchangeable errors, the residual test is asymptotically valid also for the limit hypothesis~\eqref{eq:limit_hypothesis} under the same conditions as those established in prior literature~\citep{chung2013exact, diciccio2017robust}.

\subsection{Related work}

Historically, the idea of using invariances in statistics can be traced back to~\citet{fisher1935design} and~\citet{pitman1937significance}. Due to their robustness, these procedures have been widely adopted across scientific domains, and were brought together under a common framework ---known as the randomization method--- in the seminal work of~\citet[Ch.15]{lehmann2006testing}. The literature on approximate randomization tests, on the other hand, is limited. The key theoretical tool to study such procedures remains Hoeffding's condition defined in Equation~\eqref{eq:hoeff}; see also~\citep[Sec. 15.2.2]{lehmann2006testing} for extensions. Alternatives to Hoeffding's condition, apart from the new results established in this paper, exist but tend to be tailored to particular settings. For instance, \citet{chernozhukov2021exact} studied residual-based randomization tests in a panel regression context. They proved asymptotic validity under certain mixing conditions, but their analysis relies on a single choice of $\Gall$~(i.e., permutations) and a particular linear test statistic.
  
The linear regression examples used in Section~\ref{sec:linear} to illustrate our theory are related to important works from the bootstrap and randomization literature. To be specific, by setting $\Gall$ as either the permutation group or the orthant symmetry group, our approximate randomization tests in the regression model ``rederive", respectively, the residual bootstrap~\citep{freedman1981bootstrapping, freedman1983significance, freedman1983nonstochastic} and the wild bootstrap~\citep{wu1986jackknife}. Although our results do not propose entirely novel procedures in this regression context, they provide insights into the trade-offs induced by data invariance assumptions in place of standard i.i.d. assumptions. The canonical analysis of bootstrap procedures assumes i.i.d. data under asymptotic conditions such as convergence of the covariate matrix, $\XnT X/n$; e.g., see Assumptions (1.3) \& (1.4) by~\citet{freedman1981bootstrapping}. Our theory shows that these conditions can be significantly relaxed whenever the regression errors satisfy an invariance hypothesis. Of course, the classical bootstrap results remain applicable even if the invariance hypothesis does not hold, so our results in this paper are complementary.

In a similar vein, we contribute new results to the theory of randomization-based tests employing studentized test statistics~\citep{janssen1997studentized, chung2013exact, diciccio2017robust, canay2017randomization, wu2021randomization, fogarty2023testing}. The key idea in this literature is that, in the limit, the randomization distribution of a studentized test statistic can stochastically dominate ---or be equal to--- the statistic's sampling distribution. Then, under Hoeffding's condition, the randomization test using the studentized statistic is asymptotically valid even if the invariant hypothesis does not hold. Our paper extends these results by providing new conditions for the asymptotic validity of (approximate) randomization tests under the invariant hypothesis, while being able to leverage the existing results under the limit hypothesis.

\section{Main Results}\label{sec:main}
\subsection{A feasible procedure}

The computation of  $c_{n,\alpha}$ for the randomization test defined in Equation~\eqref{eq:phi_star} is challenging because it requires enumeration of the entire invariance set, $\Gall$. To be more practical, we will redefine our tests to use random samples, $\{G_r\}_{r=1}^m$, from $\Gall$. For such a sample, define
\begin{equation}
 \Tne = \{t_n(\eps)\}  \cup \{  t_n(G_r \eps) : r=1, \ldots, m\}\nonumber
\end{equation}
 as the corresponding set of test statistic values.
For any fixed set of real numbers $\bf S$, let
\begin{equation}
c_{n, \alpha}( {\bf S}) =  \inf\big\{ z \in\Real: \frac{1}{|{\bf S} |}\sum_{s\in {\bf S}}  \Ind\{s\le z\} \ge 1-\alpha\big\}\nonumber
\end{equation}
be the modified critical value function, and define the randomization test as follows:
\begin{equation}\label{eq:phi_star2}
\phi_n^* = \Ind\big\{T_n > c_{n, \alpha}(\Tne)\big\}.
\end{equation}
In Section 1 of the Supplementary Material, we prove that this test remains valid for any $n, m>0$ despite the randomness in $\Tne$. Of course, this test remains infeasible if $\eps$ are unobserved.  To construct the (feasible) approximate randomization test, we use the proxy variables to define
\begin{equation}\label{eq:phi2}
\phione = \Ind\big\{T_n > c_{n,\alpha}(\Tner) \big\}~, 
\end{equation}
where $ \Tner = \{ T_n \}  \cup \{  t_n(G_r \heps) : r=1, \ldots, m\}$. From now on, our goal will  be to analyze the properties of this approximate test, and to derive conditions guaranteeing that the test is asymptotically valid. Before we present these results, we first give a sketch of the main argument.

\subsection{Sketch of main argument }\label{sec:sketch}

In the schematic shown below we depict the values of the original randomization test using the true variables, $\{ t_n(G_r\eps) \}$, along with the values from the approximate test, $\{  t_n(G_r\heps) \} $. 
The terms $|t_n(G_r\heps) - t_n(G_r\eps)|$  capture the {\em approximation error} from using the proxy variables rather than the true variables, while $|t_n(G_{r'}\eps) - t_n(G_{r}\eps)|$  capture the {\em spacings} in the randomization distribution of the original test. These quantities can help describe the size distortion in the approximate randomization test.

\begin{figure}[h!]
\centering
\begin{tikzpicture}[scale=0.95, transform shape]
    \pgfmathsetmacro{\ri}{
           -0.75
     }
    \pgfmathsetmacro{\rii}{0.8}
    \pgfmathsetmacro{\riii}{ 1.3 }

    \draw (0, 0) -- (10, 0);
    
          
 \draw[color=white] (7, -0.2) -- (7, 0.2) node[above] (Q) {$t_n(G_3\eps)$};
 
\node at (-1.8,0.4) {Approximate};
\node at (-1.8,-0.4) {Original};

    \foreach \x/\label/\lab in {1/$t_n(G_1\eps)$/A, 3/$t_n(G_2\eps)$/B, 7/$t_n(G_3\eps)$/C}
        \draw (\x, 0.2) -- (\x, -0.2) node[below] (\lab) {\label};

            \foreach \x/\label/\pos/\lab in {1/$t_n(G_1\heps)$/\ri/D, 3/$t_n(G_2\heps)$/\rii/E, 7/$t_n(G_3\heps)$/\riii/F} {
                \draw (\x+\pos, -0.20) -- (\x+\pos, 0.2) node[above] (\lab) {\label};
               }
                       \draw (5.5, 0.2) -- (5.5, -0.2);
                     \node at (5.5,0.5) {$T_n$};
                                              \draw (5.5, 0.25) -- (5.5, -0.25) node[below] { $t_n(\eps)$ };        
       
  \draw[thick, decorate, decoration={brace, mirror}] (A.south) -- (B.south) node[midway, below=5pt] {Spacing ($\Lambda_n$)};
  \draw[thick, decorate, decoration={brace}] (Q.north) -- (F.north) node[midway, above=5pt] {Approximation error ($\Delta_n$)};

\end{tikzpicture}
\end{figure}

The key observation in our argument is that the decision of the approximate test is, by definition, identical to the decision of the original test whenever the maximum approximation error is smaller than the minimum spacing. 
That is, $
  \phi_n = \phi_n^*~~\text{if}~\max_r |t_n(G_r\heps)-t_n(G_r\eps)| < \min_{r\neq r'} |t_n(G_r\eps) - t_n(G_{r'}\eps)|$.
The proof of Theorem~\ref{theorem:zero} then connects  these extrema to the random variables appearing 
in condition~\eqref{eq:C1}:
\begin{equation}
\Delta_n = t_n(G\heps) - t_n(G\eps),~~\Lambda_n = t_n(G'\eps) - t_n(G''\eps).\nonumber
 \end{equation}
 
Intuitively, the approximate randomization test behaves asymptotically like the original test as long as the variation in the spacings of the original test (captured by $\Lambda_n$) dominates the level of ``corruption'' introduced from using the proxy variables~(captured by $\Delta_n$).  This is the essence of condition~\eqref{eq:C1}, which we discussed earlier. Crucially, this argument is non-asymptotic and does not require convergence in distribution for either $\Lambda_n$, or $\Delta_n$, or the test statistic.  This is also the key factor that differentiates our analysis from Hoeffding's condition in Equation~\eqref{eq:hoeff}, which requires convergence in distribution of $t_n(G\eps)$.  We will also demonstrate the difference through an applied example later in Section~\ref{sec:hoeff_compare}.

\subsection{Assumptions and main result} 

Throughout our analysis, we will require the following two assumptions.
\begin{equation} 
\label{A:Gall}
\mE[t_n^2(\eps)] < \infty~\text{and}~1/|\Gall | = o(1).\tag{A1}
\end{equation}
\begin{equation}\label{A:gon}
\gon = \max_{\gg, \gg'\in\Gall, \gg\neq \gg'} P\big\{ t_n(\gg\eps) = t_n(\gg'\eps) \big\}=o(1).~\tag{A2}
\end{equation}

Assumption~\eqref{A:Gall} implies that the second moment of the test statistic exists. This moment may not converge, however, or may even be asymptotically unbounded. The second part of this assumption is also mild since   $|\Gall|$ typically increases exponentially in $n$. Assumption~\eqref{A:gon} precludes pathological cases where the test statistic becomes degenerate. To see how~\eqref{A:gon} can fail, suppose that $\eps = \eps_0 \ones_n$ ---i.e., all random components are identical--- and $\Gall$ denotes permutations~(we let ``$\ones_n$'' denote the column vector of $n$ ones). Then, $\gg \eps = \eps_0 \gg\ones_n = \eps_0\ones_n = \eps$, for any $\gg\in\Gall$. The randomization distribution is therefore degenerate. Barring such cases,~\eqref{A:gon} is satisfied with $\gon = 0$ if $t_n(\eps)$ is a continuous function.

 Our analysis will also require that the standardized spacings variable, $\LamStud= \Lambda_n/ \mVar^{1/2}(\Lambda_n)$, has a cdf with no strong discontinuities at zero. In particular, we assume that for any $n>0$ and any sequence of positive real numbers $\epsilon_n\to 0$,
 \begin{equation} 
 \label{A:degen}
 P(|\LamStud| > \epsilon_n) \to 1. ~\tag{A3}
 \end{equation}
This assumption is generally mild and aims to preclude degenerate definitions of the test statistic. For example, the assumption holds true in regular settings where $t_n(G\eps)$ converges in distribution, as in Hoeffding's condition~\eqref{eq:hoeff}. It should be noted that~\eqref{A:Gall} and~\eqref{A:gon} do not imply~\eqref{A:degen} because these two assumptions do not preclude cases where the probability density of $\LamStud$ concentrates in an ever-decreasing band around 0. Although such cases are pathological, an assumption like~\eqref{A:degen} is needed to ensure that the multiplicity in the test statistic values is controlled.

With these three assumptions in place, Condition~\eqref{eq:C1} is key for the asymptotic performance of the approximate randomization test. 
We may re-write this condition as follows:
\begin{align}
 \frac{ \mE[(t_n(G\heps) - t_n(G\eps))^2] }{  \mE[(t_n(G'\eps) - t_n(G''\eps))^2]  } 
& =  \zeta_n^2 \to 0,~\text{where}~G, G', G''\sim\Unif(\Gall)~\text{i.i.d.}.\nonumber
\end{align}
This condition formalizes the intuition built in the sketch of the main argument presented above. It does not require a particular asymptotic behavior of $\heps$ with respect to $\eps$, but only stipulates that the variation in the error from the feasible test is dominated by the variation in the original test. As an aside, notice that Condition~\eqref{eq:C1} remains invariant to certain deterministic transformations of the test statistic; e.g., $t'_n(\eps) = a_n(\eps)[t_n(\eps) + b_n]$, where $a_n(\gg\eps)=a_n(\eps)\in\mathbb{R}^+$, for all $\gg\in\Gall$. This is appropriate because such transformations of the test statistic do not change the decision of the randomization test. We are now ready to state our first main result.

\begin{theorem}
\label{theorem:zero}
\TheoremZero
\end{theorem}
\begin{remark}\label{rem1}
(a) The theorem shows that $\phi_n$ and $\phi_n^*$ are asymptotically equivalent. Under the invariant hypothesis~\eqref{eq:invariant_hypothesis}, this immediately implies the asymptotic validity of $\phi_n$, since that hypothesis implies $\mE(\phi_n^*)\le \alpha$ for any $n>0$. If the invariance property does not hold true, then $\phi_n^*$ is asymptotically valid as long as $\phi_n^*$ is itself asymptotically valid (e.g., via studentization). 

(b) The $m^{-1}$ term in the expression of Theorem~\ref{theorem:zero} exists because we allow for multiplicities in the test statistic values. Indeed, the term vanishes if the test statistic values are unique with probability 1. In some cases,  we could correct the decision by $\alpha / [\alpha + (m+1)^{-1}] = 1-O(1/m)$ to obtain a level $\alpha$ test.
\end{remark}

\begin{example}[Independence test]
\label{example1}
\singlespacing
\em
To illustrate Theorem~\ref{theorem:zero}, consider two random variables, $\eps = (\eps_1, \ldots, \eps_n)$ and $u = (u_1, \ldots, u_n) \in \Real^n$, with finite second moments.
Suppose we want to test $H_0: \eps ~\indep~ u $, under the invariance assumption that $\eps$ are exchangeable. 
That is, Equation~\eqref{eq:invariant_hypothesis} holds for $\eps$
with $\Gall$ as the group of $n\times n$ permutation matrices---denote this group by $\Gall^\pp$.

For simplicity, we use the inner product as the test statistic,
$$
T_n = \sum_{i=1}^n (\eps_i - \bar\eps)(u_i - \bar u)  :=  \tilde\eps' \tilde u~,
$$
where,  for any given vector $z\in\Real^n$, $\tilde z= (z_1-\bar z, \ldots, z_n-\bar z)$ denotes the centered version of $z$. If both $\eps$ and $u$ can be fully observed, then the randomization test defined in~\eqref{eq:phi_star2} using $T_n$ as the test statistic is finite-sample valid in the sense that $\mE(\phi_n^*) \le \alpha$ whenever $H_0$ is true.

Suppose, however, that $\eps$ can only be observed with noise 
through the proxy variables $\heps_i = \eps_i + \xi_i$, where 
$\xi = (\xi_1, \ldots, \xi_n)$ is allowed to depend on either $\eps$ or $u$ in an arbitrary way. Under what conditions is the approximate test based on $\heps$ asymptotically valid? 

To answer this question, first define $t_n(\eps) = \tilde\eps' \tilde u$, in the hypothetical case where all variables can be observed.
For $G\sim\Unif(\Gall^\pp)$, and any fixed $z \in \Real^n$, we use
\begin{equation}\label{eq:identity_perm}
 \mE( G \tilde z \tilde z' G^\top )=\mE( G^\top \tilde z \tilde z' G ) = s^2_z \Id_n,
\end{equation}
where $s^2_z= 1/(n-1)\sum_{i=1}^n (z_i-\bar z)^2$ is the spread of $z$, 
and $\Id_n$ is the $n\times n$ identity matrix. We prove this result in the Supplementary Material as part of the proof for Theorem~\ref{thm:perm}.
Moreover, under the null hypothesis, 
$\mE[ t_n(G \eps) | \eps ] =  \tilde \eps' \mE(G \tilde u)  = \tilde\eps'\overline{\tilde u} = 0$. The exchangeability of $\eps_i$ then implies 
$\mE[(t_n(G'\eps) - t_n(G''\eps))^2]= 2\mE[t_n^2(G\eps)]$ and so
$$
\mE[(t_n(G'\eps) - t_n(G''\eps))^2]= 2 \mE( \tilde u' G \tilde{\eps} \tilde \eps' G^\top \tilde \eps ) = 2(n-1) \mE(s^2_\eps s^2_u) = 2(n-1) \mE(s^2_\eps) \mE(s^2_u).
$$
Here, the last equality follows from  independence of $\eps$ and $u$ under the null. From independence of $G$, result~\eqref{eq:identity_perm} also implies
$$
\mE[ (t_n(G\heps) - t_n(G\eps) )^2 ] = \mE( \tilde u' G \tilde \xi \tilde \xi' G^\top \tilde u ) =  
(n-1) \mE(s^2_\xi s^2_u).
$$
Putting these together, we can reduce Condition~\eqref{eq:C1} to
\begin{equation}\label{eq:example_ratio}
 \frac{  \mE(s^2_\xi s^2_u)   } { \mE(s^2_\eps) \mE(s^2_u) } \to 0.
\end{equation}

The approximate randomization test is therefore valid asymptotically whenever the variation in $\eps$ dominates the variation in the noise, $\xi$. Importantly, this condition does not require that $\eps_i, u_i$ or $\xi_i$ are identically distributed, or that they have converging or even  bounded moments. In contrast, Hoeffding's condition in Equation~\eqref{eq:hoeff} posits that $t_n(G\heps)$ ---or an appropriate transformation of it--- converges in distribution.  This stronger requirement as compared to~\eqref{eq:example_ratio} is not too surprising since Hoeffding's condition does not assume that $\eps_i$ are invariant. Relating to the discussion in the Introduction, this example highlights how our theory addresses existing theoretical gaps in our understanding of approximate randomization tests under the invariant hypothesis.
\end{example}

\subsection{Numerical illustration}\label{sec:example1}
Here, we illustrate the analysis of Example 1 through a simulation study. Let
\begin{equation}\label{eq:example}
\eps_i \sim \mathrm{N}(0, 1),~u_i \sim \mathrm{N}(0, 1),~\xi_i = n^{-b} \big[\rho_n u_i + \sqrt{1-\rho_n^2} \mathrm{N}(0,1)\big].
\end{equation}
Thus, $\rho_n$ controls the dependence between $\xi_i, u_i$ while $b$ controls the ``signal-to-noise'' ratio. The ratio in Equation~\eqref{eq:example_ratio} is then approximately equal to $n^{-2b}$. Larger values of $b$ therefore indicate larger signal-to-noise, which we expect to be beneficial for the approximate randomization test according to the preceding analysis. In the simulation study, we consider $\rho_n = 1/\sqrt{n}$ or $\rho_n=0.1$, and  $b \in \{0, 0.2, 0.5, 0.9\}$. Note that the setting with $\rho_n=0$ is trivial because it implies independence between $\xi$ and $u$, and thus the finite-sample validity of the approximate test. 

The results for $\rho_n=1/\sqrt{n}$ over 500,000 replications are shown in Table~\ref{tab1}. For $b=0$, we observe that the approximate randomization test does not achieve the correct level asymptotically (as $n$ grows). This is expected from our theory since the ``noise-to-signal'' ratio in~Equation~\eqref{eq:example_ratio} fails to reach 0 when $b=0$. To confirm, see the last column of the table, which reports the ratio averaged over all replications, and specifically the sub-panel corresponding to $b=0$. We see that the ratio in that sub-panel fails to reach 0. However, when $b>0$, we see that the approximate test attains the correct level asymptotically as $n$ grows and the ratio~in~\eqref{eq:example_ratio} converges to 0. The results for $\rho_n=0.1$ are similar, with the only difference being that the approximate test tends to over-reject more when $b=0$ due to the stronger correlation between $\xi$ and $u$.

\begin{table}
\centering
	\begin{tabular}{rrrrr}
  \hline
$b$ & $n$ & True test & Approximate test & Ratio in~Eq.~\eqref{eq:example_ratio} \\ 
 \hline
   0 & 50 & 4.96 & 10.56 & 1.00 \\ 
  0 & 100 & 5.00 & 10.72 & 1.00 \\ 
  0 & 200 & 4.98 & 10.78 & 1.00 \\ 
  0 & 500 & 5.01 & 10.81 & 1.00 \\ 
    \hline
  0.20 & 50 & 4.99 & 6.86 & 0.21 \\ 
  0.20 & 100 & 5.02 & 6.52 & 0.16 \\ 
  0.20 & 200 & 5.01 & 6.17 & 0.12 \\ 
  0.20 & 500 & 4.96 & 5.81 & 0.08 \\
     \hline 
  0.50 & 50 & 5.02 & 5.23 & 0.02 \\ 
  0.50 & 100 & 5.06 & 5.14 & 0.01 \\ 
  0.50 & 200 & 5.01 & 5.06 & 0.01 \\ 
  0.50 & 500 & 5.00 & 5.00 & 0 \\ 
      \hline
  0.90 & 50 & 5.02 & 5.01 & 0 \\ 
  0.90 & 100 & 5.02 & 5.02 & 0 \\ 
  0.90 & 200 & 4.98 & 4.96 & 0 \\ 
  0.90 & 500 & 5.09 & 5.10 & 0 \\ 
      \hline
\end{tabular}
	\label{tab1}
	\begin{caption}
Results from the simulation of Example 1.
	 `True' and `Approximate' represent, respectively, the rejection rates (\%) for $\phi_n^*$ and $\phi_n$ for the simulation study of Section~\ref{sec:example1}. The nominal level is set at 5\% for both tests.
	\end{caption}
\end{table}

\begin{example}[continued]
\singlespacing
\em
In Example~\ref{example1}, suppose that $\eps_i$ are not exchangeable but they are symmetric around 0. That is, invariance~\eqref{eq:invariant_hypothesis} holds under the group of $n\times n$ diagonal matrices with $\pm 1$ in the diagonal. Let $\Gall^{\ss}$ denote this group of transformations, which we will refer to as the orthant symmetry group~\citep{efron1969student}. Under this invariance, the analysis of the independence test of Example~\ref{example1} can follow the same steps as the original analysis under exchangeability, requiring us only to update~Equation~\eqref{eq:identity_perm} to
\begin{equation}\label{eq:identity_sign}
\mE( G\tilde z \tilde z' G^\top ) = \diag_n(\tilde z_i^2),~~G\sim\Unif(\Gall^{\ss}).
\end{equation}
Here, $\diag_n(u_i^2)$ is the $n\times n$ diagonal matrix formed by the elements $u_i^2$.  See the Supplementary Material for a proof.
Condition~\eqref{eq:C1} is then simplified to
$$
\frac{  \mE[ \tilde u' \diag_n(\tilde \xi_i^2) \tilde u ]  } {  \mE[ \tilde u' \diag_n(\tilde \eps_i^2) \tilde u ]  } 
\le  \frac{  \mE[ \xi_{(n)}^2  s^2_u ] } {  \mE[ \eps_{(1)}^2] \mE(s^2_u)  }  \to 0.
$$
Here, $x_{(i)}$ denotes the $i$-th order statistic in a sample $\{x_1, \ldots, x_n\}$. The asymptotic validity condition now depends on the extrema of the distributions of $\eps_i$ and $\xi_i$, and is somewhat stricter than the validity condition under exchangeability derived in Equation~\eqref{eq:example_ratio}. Intuitively, each individual observation has more ``leverage" in a randomization test using random sign flips than in a randomization test using random permutations.
\end{example}

\section{Non-asymptotic analysis}
\label{sec:finite_sample}
In this section, we derive a non-asymptotic bound for the difference between the test sizes of  $\phi_n$ and $\phi_n^*$. This can help us understand the finite-sample properties of approximate randomization tests, including the residual-based tests for linear regression described in the following section.

To proceed, we need to introduce 
a regularity condition on how the standardized spacings variable, $\LamStud = \Lambda_n/ \mVar^{1/2}(\Lambda_n)$, behaves near 0.
In particular, we impose a~$\gamma$-H\"{o}lder continuity property on $F_{\LamStud}$,  the cdf of $\LamStud$, and assume there exists $\gamma>0$ such that for any $n>0$ and any sufficiently small $\epsilon>0$,
\begin{align}
\label{A:gammas}
\FLn(\epsilon) - \FLn(-\epsilon) = O(\epsilon^\gamma) +  O(\gon + 1/|\Gall|).~\tag{A3'}
\end{align}

This assumption will supersede~\eqref{A:degen} because~Assumptions~\eqref{A:Gall}-\eqref{A:gon} and~\eqref{A:gammas} imply~\eqref{A:degen}. Roughly speaking, parameter $\gamma$ is associated with the tails of the normalized spacings.  For instance, if $\LamStud$ is a Weibull random variable, then values of $\gamma$ less than 1 correspond to heavier tails. If $\LamStud$ is asymptotically normal, then $\gamma$ is approximately 1. For more intuition, see the Supplementary Material for a simulation where we numerically estimate $\gamma$ under various settings.
\begin{theorem}
\label{thm:main}
\TheoremOne
\end{theorem}

Theorem~\ref{thm:main} provides insights into the finite-sample performance of approximate randomization tests. The performance depends on the multiplicity of test statistic values ($\gon$),  the invariance structure $(\Gall)$, and the smoothness in the  distribution function of the spacings variable ($\gamma$).

The theorem also shows that the rate at which Condition~\eqref{eq:C1} is satisfied determines the non-asymptotic bound, $\zeta_n^{2\gamma/(2+\gamma)}$, on the discrepancy between the approximate test and the true test. Notably, this bound is reminiscent of the minimax convergence rate in non-parametric regression~\citep[Section 1.6]{tsybakov2009}.
Under standard regularity conditions, $\LamStud$ is asymptotically normal 
and so $\zeta_n^2=1/n$ and $\gamma=1$, implying the rate $O(n^{-1/3})$.  Heavier-tailed data may ``slow down'' this asymptotic rate even more. If, in addition, the invariance hypothesis holds~\eqref{eq:invariant_hypothesis}, then Theorem~\ref{thm:main} implies a finite-sample bound on the Type I error of the approximate test:
$$
\mE(\phione)  \le \alpha+ O(1/m) + O(m\gon) + O(m^2/ |\Gall|)  +  A(\gamma, m)\cdot O(n^{-1/3}).
$$

Next, we use Theorem~\ref{thm:main} to establish a finite-sample bound for the Type II error of the approximate randomization test. This shows that the approximate test is consistent as long as the ``signal'' dominates the natural variation in the true randomization test.
\begin{theorem}\label{thm:power}
\TheoremPower
\end{theorem}
\begin{remark}\label{rem:power}
(a) The result in~Theorem~\eqref{thm:power} is in line with~\citet{dobriban2021consistency}, who also studied the consistency of invariance-based randomization tests. The consistency results in that paper, however, focus on exact randomization tests in a ``signal plus noise'' regime, while our focus here is on approximate tests under an arbitrary invariance structure.

(b) A simple way to satisfy $P(|t_n(G\eps) - t_n(\eps)| >  \tau_n/\sigma_n)\to0$ is if $t_n(\eps), t_n(G\eps)$ converge in distribution, but not necessarily according to the same law. This is true, for example, under Hoeffding's condition~\eqref{eq:hoeff}.

(c) In the sections that follow, we will apply our theory in a linear regression setting by using approximate randomization tests for significance testing on regression coefficients. Under standard OLS assumptions, the power expression in~Theorem~\ref{thm:power} simplifies to $1-e^{- c \tau_n^2} - O(n^{-1/3})$, where $\tau_n$ is the local alternative (signal strength). See the Supplementary Material for more details.
\end{remark}

\begin{example}[Median of a symmetric distribution]
\singlespacing
\em
Here, we revisit a classical example that tests the symmetry of a random variable around 0; see \citep[Section 4]{hoeffding1952},~\citep[Theorem 15.2.4]{lehmann2006testing}. Let $\eps = (\eps_1, \ldots, \eps_n)$ be random variables, and consider a null hypothesis implying that each $\eps_i$ is symmetric around zero. The null hypothesis is therefore equivalent to the invariant hypothesis~\eqref{eq:invariant_hypothesis} under the orthant symmetry group, $\Gall^\ss$.

In this context, the classical analysis considers the test statistic $T_n' = \sum_i \eps_i / (\sum_i \eps_i^2)^{1/2}$. If $\eps_i$ are independent and identically distributed with finite second moments, then Hoeffding's condition in Equation~\eqref{eq:hoeff} holds true. This implies that the test is valid ---validity holds in finite samples in this case--- and, importantly, the test has power 1 in the limit. In fact, in this setting, the randomization test has asymptotically the same power as the standard $t$-test~\citep{hoeffding1952}.

Our theory can complement these classical results. To make the problem a bit more challenging,  suppose that $\eps_i$ are unobservables, but proxies $\heps_i = \eps_i + \xi_i$ are available. To keep the analysis simple, let us also define the test statistic based on the sample mean:
$T_n = (1/\sqrt{n})\sum_i \heps_i := t_n(\heps)$. Then, for $G\sim\Unif(\Gall^\ss)$,
\begin{equation}
\mE[(t_n(G\heps) - t_n(G\eps))^2] =\mE(\ones_n' G \xi \xi' G^\top \ones_n) /n
 = \mE[\ones' \diag_n(\xi_i^2) \ones_n] /n= E(||\xi||^2/n). \nonumber
\end{equation}
Here, in the second equality we used the result from Equation~\eqref{eq:identity_sign}. Similarly, we obtain
$$
E[(t_n(G\eps) - t_n(G'\eps))^2] = 2 E[t_n^2(G\eps)]/n = 
2 E(\ones_n' G \eps \eps' G^\top \ones_n)/n = 2 E(||\eps||^2/n).
$$
\singlespacing
Thus, condition~\eqref{eq:C1} simplifies to an intuitive ``signal-to-noise" ratio condition:
$$
\frac{E(||\xi||^2)}{E(||\eps||^2)} \to 0.
$$
If this condition holds, and the sample mean statistic is not degenerate in the sense of Assumption~\eqref{A:degen}, then Theorem~\ref{theorem:zero} implies that the randomized sign test using proxies $\heps$ is asymptotically valid for the null hypothesis that $\eps_i$ are symmetric around 0. Similar to the previous example, note that this condition does not impose any further restrictions on the joint distribution of $(\xi_i, \eps_i)$ beyond their marginal second moments and the invariance on $\eps_i$.

Furthermore, consider an alternative space where $\bar \eps = \mu_n >0$ and $\eps_i, \xi_i$ are independent random variables with mean zero and finite second moments, such that $t_n(\eps)$ and $t_n(G\eps)$ both converge in distribution. Then, Theorem~\ref{thm:power} implies that the randomized sign test has power 1 asymptotically as long as $ \sqrt{n} \mu_n / [\mE(||\eps||^2/n)]^{1/2} \to 0$.
\end{example}

\section{Examples from Linear Regression}\label{sec:linear}

\subsection{Setup}
In this section, we apply our theory to tests of significance in linear regression. The underlying model is defined as 
\begin{equation}\label{eq:linear_model}
y_i =  x_i'\beta + \eps_i,
\end{equation}
where $(y_1, x_1), \ldots, (y_n, x_n) \in \Real \times \Real^p$ are observed data, $\beta = (\beta_0, \ldots, \beta_{p-1})\in \Real^p$ are fixed parameters, and $\eps_i$ are unobserved errors; $x_i$ has always one as the first element; $\beta_0$ is the intercept. Let $X = (x_1, \ldots, x_n)$, $ y = (y_1, \ldots, y_n)$ and $\eps = (\eps_1, \ldots, \eps_n)$. 
Also let $\Pn = X (\XnT X)^{-1} \XnT$ be the projection matrix in the column space of $X$. Our procedures will rely on regular least squares estimators, so we need to update Assumption~\eqref{A:Gall} to make sure that these estimators are well-defined. We will therefore assume that
\begin{flalign} 
\label{A:X_eps2}
\lmin(\XnT X)>0,~\text{and}~\mE(\eps\eps') < \infty,~
\text{and}~1/|\Gall| = o(1).\tag{A1'}
\end{flalign}
Our focus will be to test hypotheses of significance on individual coefficients:
\begin{equation}\label{H0_reg}
H_0^j:  \beta_j= 0,~j\in\{1, \ldots, p\}.
\end{equation}

To illustrate the various aspects of our theory, we will test $H_0^j$ 
using two different test statistics. One test statistic is simple and is defined as $T_n = e_j' \hbeta$, where $e_j\in\{0,1\}^n$ is one only at the $j$-th component and $\hbeta=(\XnT X)^{-1} \XnT y$ is the least squares estimator. We will also consider the studentized statistic: $\Tstud = e_j'\hbeta/ \sigma_n(\heps)$, where $\sigma_n^2(u) = e_j'(\XnT X)^{-1} X^\top \diag_n(u_i^2) X (\XnT X)^{-1} e_j$, and $\heps_i = y_i - x_i'\hbeta$ denotes the regression residuals. Under $H_0^j$, the unstudentized test statistic and the studentized test statistic can be described, respectively, by the functions:
\begin{equation}\label{eq:linear_Tn}
T_n\myeq{H_0^j} e_j' (\XnT X)^{-1} \XnT \eps :=t_n(\eps),~\text{and}~
T_n^s \myeq{H_0^j} \frac{t_n(\eps)}{\sigma_n(\eps - \Pn\eps)} := t_n^s(\eps).
\end{equation}

In relation to these test statistics, we will consider both a class of invariant hypotheses and a class of limit hypotheses. The purpose of these  hypotheses is to impose a structure in the distribution of $(X, \eps)$ in order to study the properties of our randomization tests. First, for a given choice of $\Gall$, we will consider invariant hypotheses of the following form:
\begin{equation}\label{eq:invariant_hypothesis_error}
    \eps \myeq{d} \gg \eps \mid X,~~\text{for all}~X,~\gg\in\Unif(\Gall).\tag{Inv}
\end{equation}
This assumption can express exchangeable or sign symmetric errors conditional on covariates $X$. Which assumption is more plausible depends on the application. For example, sign symmetry may be more appropriate than exchangeability in settings with heteroskedasticity.

Alternatively, in place of invariance assumptions, we can make all standard assumptions of ordinary least squares regression~\citep[Chapter 4]{wooldridge2010econometric}; i.e.,
\begin{equation}\label{eq:limit_hypothesis_error}
\text{i.i.d.} \{(x_i, \eps_i)\}_{i=1}^n,~\mE(\eps_i |x_i) = 0,~\text{non-singular}~\mVar(x_i),~ 
\mVar(\eps_i x_i),~\mE(\eps_i^4)<\infty,~\mE(x_{ij}^4) < \infty,~j\in[p].\tag{Lim}
\end{equation}
We will collectively refer to these assumptions as the limit hypothesis~\eqref{eq:limit_hypothesis_error}. Under these assumptions, prior work has established that both the sampling distribution of the studentized test statistic, $t_n^s(\eps)$, and its randomization distribution, $t_n^s(G\eps)$, are asymptotically standard normal under certain invariances~\citep{janssen1999nonparametric, diciccio2017robust}. We will leverage these results to study our residual-based randomization tests for $H_0^j$. We describe these tests next.

 \subsection{Residual randomization tests of significance}

Since $\eps$ are unobserved,  we will make use of the least squares residuals as proxy variables. For a given invariance $\Gall$, the residual randomization distribution is defined as
$$
\mathbf S_n = \{ T_n \}  \cup \{  t_n(G_r \heps)\}_{r=1}^m
$$
where $G_r\sim\Unif(\Gall)$ are randomly sampled transformations from $\Gall$.
Thus, we can define the residual randomization test under $\Gall$ as follows:
\begin{equation}\label{eq:phi_linear}
\phi_n = \Ind\big\{T_n > c_{n,\alpha}(\mathbf S_n ) \big\}.
\end{equation}
The definitions for the studentized statistic are completely analogous:
$\mathbf S_n^{\mathrm{s}} = \{ \Tstud \}  \cup \{  t_n^s(G_r \heps)\}$, 
and $\phi_n^s = \Ind\{\Tstud > c_{n,\alpha}(\mathbf S_n^s )\}$.
For both of these residual-based tests, another possibility is to use restricted residuals, calculated under the least squares estimate constrained by the null hypothesis. In the proofs, we analyze both regular residuals and restricted residuals, showing that these two approaches are in fact asymptotically equivalent. 

It is important to note that these residual randomization tests are related to well-known bootstrap variants under certain invariances, a point also made by~\citet{diciccio2017robust}. Specifically, when we  set $\Gall$ as the permutation group, the resulting residual permutation test may be viewed as the randomization analogue to the celebrated residual bootstrap method~\citep{freedman1981bootstrapping, freedman1983significance, freedman1983nonstochastic}. The two procedures differ only on whether sampling is with or without replacement.  Alternatively, by setting $\Gall$ as the orthant symmetry group (i.e., sign flips),  the resulting residual sign test becomes the randomization analogue to the wild bootstrap~\citep{wu1986jackknife}. Consequently, the theory developed in the following sections provides a complementary set of conditions for the validity of bootstrap procedures under error invariance assumptions such as~\eqref{eq:invariant_hypothesis_error}.  This complements the asymptotic equivalence results between bootstrap and randomization tests established in the seminal work of~\citet{romano1989bootstrap}.

\subsection{Residual permutation tests}
\label{sec:perm}

We begin with $\Gall$ defined as the permutation group $\Gall^\pp$. 
This results in ~$\phi_n$ being a residual permutation test. 
Without loss of generality, we will also assume that the regression model is centered such that $\XnT \ones_n=0$ and $\bar\eps=0$.
Operationally, the residual permutation test compares the observed value of the test statistic with values of the statistic calculated at randomly permuted residuals. This leads to a randomization-based analogue of the residual bootstrap procedure of~\citet{freedman1981bootstrapping}. However, the validity conditions for the 
residual permutation test vary substantially depending on the hypothesis regime, as shown in the theorem below.
\begin{theorem}\label{thm:perm}
\TheoremPerm
\end{theorem}

Theorem~\ref{thm:perm} has a dual interpretation. First, if $\eps$ are exchangeable conditional on $X$, the theorem shows that the residual permutation test is asymptotically valid under a remarkably simple condition: $p/n=o(1)$. This allows covariates with high leverage or outliers, and even certain high-dimensional cases where $p <n$ but $p \to\infty$.  If, in addition, Assumption~\eqref{A:gammas} holds, then the Type I error of $\phi_n$ can be bounded using Theorem~\ref{thm:main} as follows:
\begin{equation}\label{eq:perm_bound}
\mE(\phi_n) \le \alpha + O\big( (p/n)^{\gamma/(2+\gamma)} \big).
\end{equation}
The rate of convergence slows down whenever $p$ grows fast with respect to $n$. It is remarkable, however, that $p/n$  is the only important factor for the convergence rate, and no further restrictions are imposed on the behavior of $X,\eps$ beyond what is already assumed in~\eqref{A:Gall}-\eqref{A:gammas}. This shows that exchangeability is a strong assumption, in line with the concept of 
exchangeability as a ``mixture of i.i.d. sequences'' from de Finetti's theorem.

A second implication of the theorem relates to the limit hypothesis. In particular, under~\eqref{eq:limit_hypothesis_error}, the theorem states that the residual permutation test based on the studentized statistic, $t_n^s(G\heps)$, has asymptotically the same size as the idealized permutation test based on the true errors, $t_n^s(G \eps)$. Since the asymptotic validity of the idealized test has already been established in prior work~\citep[Theorem 3.3]{diciccio2017robust}, it follows that the approximate test is asymptotically valid as well. From this perspective, our theory unifies the analysis across the two hypothesis regimes: it provides simple conditions for asymptotic validity under the invariant hypothesis, while still being able to leverage existing results under the limit hypothesis.

\begin{remark}
A similar duality of interpretation exists in relation to bootstrap theory. For instance, in the analysis of the residual bootstrap procedure, Assumptions (A1.3) \& (A1.4) in~\citep{freedman1981bootstrapping} require i.i.d.~errors with constant variance and a limit for $\XnT X/n$. See also~\citep{bickel1981some, lopes2014residual} for the high-dimensional regime. Our analysis shows that these conditions are not necessary if the errors are exchangeable conditionally on $X$. It should be noted that no set of assumptions, whether invariance-based or i.i.d., is a special case of the other. They are simply two different kinds of theoretical assumptions, 
each leading to unique insights into the behavior of resampling-based procedures such as the randomization test and the bootstrap.
\end{remark}

\begin{example}[Simulated study under exchangeability]
\label{example4}
\singlespacing
\em
To illustrate the result of Theorem~\ref{thm:perm}, consider the linear model $ y = X\beta + \eps$ in~\eqref{eq:linear_model} with $n \in \{50, 100, 250, 500, 800\}$ and $p \in \{5, \log n, n^{1/3}, n^{1/2} \}$. The errors are sampled as $\eps_i \sim N(0,1)$ or $\eps_i \sim \mathrm{t}_3$, all i.i.d; The covariates are sampled as $x_{ij} \sim \mathrm{Weibull}(1)$ or $x_{ij} \sim \mathrm{Weibull}(1)$, $i=1, \ldots, n$, $j=1, \ldots, p$, i.i.d. We set the true parameters as $\beta = (-1, 0, 0, \ldots)$, and test $H_0: \beta_1=0$. In each simulation setting, we apply the residual permutation test as described by Equation~\eqref{eq:phi_linear} with $m=1,000$.

The results of this simulation are shown in Table~\ref{tab:perm} and are split into two scenarios: One regular scenario with normal errors (left panel) and another scenario with heavy-tailed errors and covariates (right panel). Note that this simulation also considers high-dimensional settings with diverging $p$. The table reports rejection rates (\%)  estimated over 500,000 replications with the nominal level set at 5\%. We see that, across all settings, the rejection rates of the residual permutation test converge to the nominal level of 5\% for any fixed $p$. We can also see that the scenario with heavy tails does not seem to make any difference on  convergence to the nominal level. The residual test remains robust across scenarios.

What seems to matter most for the convergence of the Type I error rate towards the nominal level as $p,n\to\infty$ is the problem dimension, $p$. For instance, we can see that the convergence towards the 5\% level is noticeably faster when $p=5$ compared to $p=n^{1/2}$. This holds true in both the regular and the heavy-tailed scenario. This observation can be seen as empirical validation for the theoretical bound on the test's Type I error rate derived in Equation~\eqref{eq:perm_bound}, which was based on  Theorem~\ref{thm:perm} combined with the finite-sample result of Theorem~\ref{thm:main}. Empirically, the (log) over-rejection from Table~\ref{tab:perm} in relation to $\log(p/n)$ is indeed linear, exactly as predicted by the finite-sample bound in Equation~\eqref{eq:perm_bound}. A linear regression fit (not shown here) of $\log(|\mE(\phi_n)-\alpha|)$ on $\log(p/n)$ yields a coefficient that is slightly smaller than 1.

\renewcommand{\arraystretch}{1.1}
\begin{table}[t!]
\centering
\begin{tabular}{rrrrr | rrrr}
  \hline
  \hline
& \multicolumn{4}{c}{ $\eps_i\sim N(0,1), ~x_{ij}\sim \text{Weibull}(1)$} &  \multicolumn{4}{c}{$\eps_i\sim \mathrm{t}_3, ~x_{ij}\sim \text{Weibull}(0.5)$} \\
$n$ & \multicolumn{4}{c}{ $ p $ }  & \multicolumn{4}{c}{ $ p $ } \\
\hline
 & 5 & $\log n$ & $n^{1/3}$ & $n^{1/2}$  & 5 & $\log n$ & $n^{1/3}$  & $n^{1/2}$  \\ 
  \hline
50 & 5.77 & 5.19 & 5.22 & 6.39 & 5.63 & 5.16 & 5.22 & 6.21 \\ 
  100 & 5.35 & 5.22 & 5.24 & 6.02 & 5.33 & 5.18 & 5.20 & 5.86 \\ 
  250 & 5.13 & 5.15 & 5.24 & 5.64 & 5.16 & 5.19 & 5.18 & 5.57 \\ 
  500 & 5.05 & 5.12 & 5.11 & 5.56 & 5.00 & 5.10 & 5.07 & 5.48 \\ 
  800 & 5.03 & 5.06 & 5.08 & 5.34 & 5.06 & 5.05 & 5.07 & 5.33 \\ 
\hline
\end{tabular}
\caption{Rejection rates (\%) for the residual permutation test in the simulation of Example~\ref{example4}. 
}
\label{tab:perm}
\end{table}
\end{example}

We conclude this section with an example, originally discussed in~\citep[Section 6]{hoeffding1952}, comparing our theory with Hoeffding's condition in Equation~\eqref{eq:hoeff}. In our adaptation of this example, Hoeffding's condition fails but the approximate
randomization test remains asymptotically valid in line with our theory. 

\subsection{Comparison to Hoeffding's condition}\label{sec:hoeff_compare}
Consider a simple linear model
$$
    y_i = \beta_0 + \beta_1 x_{i} + \eps_i,
$$
where $\eps_i$ are exchangeable conditional on $x_i$, but otherwise have an arbitrary distribution. We wish to test $H_0:\beta_1=0$. The hypothesis is, of course, not that interesting ---in fact it can be tested in an exact way---  but the real objective here is to demonstrate our theory. As in the original example, we will use the $t$-statistic,
$$
T_n = \frac{\sum_i(x_i-\overline{x}) (y_i - \bar y)}{ 
\{ \sum_i (x_i-\bar x)^2 (n-1)^{-1} \sum_i (y_i-\bar y)^2\}^{1/2}}.
$$
Under the null hypothesis, this statistic can be written as
 $$
 T_n \myeq{H_0} \frac{\sum_i(x_i-\overline{x}) (\eps_i - \bar \eps)}{ 
\{ \sum_i (x_i-\bar x)^2 (n-1)^{-1} \sum_i (\eps_i-\bar \eps)^2\}^{1/2}} := t_n(\eps),
 $$
 
 In a classical result~\citep[Theorem 6.1]{hoeffding1952}, it was shown that the permutation distribution of $T_n$ satisfies Hoeffding's condition if the $x_i$s are not too highly leveraged:
\begin{equation}\label{eq:hoeff_2}
    \frac{\max_{1\le i \le n} (x_{i} - \overline{x})^2}{\sum_i (x_{i}-\overline{x})^2 } 
    \to 0.
\end{equation}
Theorem~\ref{thm:perm}, in contrast, offers a much simpler condition. Under exchangeable $\eps_i$, our theorem implies that the residual permutation test should be valid asymptotically as $n\to\infty$ without any additional requirements on the distribution of $x_i$.

To understand this discrepancy, let's construct a scenario where the covariates are highly leveraged. First, let $I_n \in [n]$ be a sequence of positive integers, $\epsilon>0$ some small positive constant, and define $s_n\to\infty$ as some diverging sequence of real numbers. Then, for any given $n>0$, define $x_{i} = \epsilon$ if $i\neq I_n$, and $x_{I_n} = s_n$. As a result, $x_{I_n}$ is a highly leveraged observation, and the ratio in Equation~\eqref{eq:hoeff_2} is equal to $n/(n-1) \to 1$, failing Hoeffding's condition. Indeed, a straightforward calculation implies
\begin{equation}\label{eq:hoeff_leveraged}
t_n(\eps) = [s_n/(s_n-\epsilon)+o(1)] (\eps_{I_n} - \bar \eps) = \eps_{I_n} - \bar\eps + o_P(1).
\end{equation}

That is, the test statistic is dominated by the highly-leveraged observation determined by $I_n$. As a result, $t_n(G\eps)$ ---where $G$ is a random permutation from $\Gall^\pp$--- does not converge in distribution since the distribution of $\eps_i$ is arbitrary. Moreover, $t_n(G\eps), t_n(G' \eps)$ are generally not independent because, for an exchangeable vector $\eps$, any two elements $\eps_{I_n}, \eps_{\gg(I_n)}$ may still be mutually dependent. In this case, standard asymptotic theory of permutation tests is not applicable.

However, even though the test statistic itself does not converge in distribution, Equation~\eqref{eq:hoeff_leveraged} reveals that the permutation test will still be valid. Indeed, $t_n(\heps) \approx \heps_{I_n}$, and so the test statistic remains invariant to permutations since, roughly speaking, $t_n(\pi\heps) \approx \heps_{\pi(I_n)} \myeq{d} \heps_{I_n} \approx t_n(\heps)$, for a random permutation $\pi$. We emphasize that with this example we do not propose a new procedure to test the significance of $\beta_1$. We only point out that the classical permutation test can be valid under weaker conditions than what is currently predicted by standard theory, if one is willing to assume exchangeable errors. The following numerical simulation illustrates this point.

\begin{example}[Simulation with high-leveraged regression]
\label{example:hoeffding}
\singlespacing
\em
Here, we will slightly complicate the example of the previous section as follows:
    $$
    y_i = -1 + 0 x_{1i} + 0.1 x_{2i} + 0.2 x_{3i} + \eps_i.
    $$
We sample covariates as $x_{2i}, x_{3i}\sim U[0,1]$ i.i.d. uniform random variables. We also sample $x_1$ in a way that leads to heavy tails and unbounded moments:  $x_{1i} = B_i C_i + (1-B_i) \sqrt{n}$, where $B_i$ are i.i.d. Bernoulli  and $C_i$ are i.i.d. Cauchy random variables.  The errors are sampled i.i.d. as $\eps_i= \lambda(X) {\tt t}_{3i}$, where $\lambda(X) = (||X||^2/np)^{1/4}$ captures the signal strength, and ${\tt t}_{3}$ denotes the ${\tt t}$ distribution with 3 degrees of freedom. We will test the null $H_0: \beta_1=0$ at the 5\% level, which holds true in our model.

The results of this simulation study using the residual permutation test are shown in Table~\ref{tab:hoeff_perm} for samples sizes $n\in\{20, 50, 100, 250\}$.  The two columns on the right aim to highlight the theoretical points that we discussed above. These columns calculate  Hoeffding's leverage quantity in Equation~\eqref{eq:hoeff_leveraged}  and our ratio in Condition~\eqref{eq:C1}, averaged across simulations. First, we see that Hoeffding's leverage quantity fails to converge to 0 in this case, remaining roughly equal to 0.6 through the simulations. This failure to converge is not surprising since $x_1$ is a heavy tailed random variable with an unbounded second moment. As discussed above, this also means that the permutation distribution of the test statistic is not normal asymptotically.

However, the ratio in Condition~\eqref{eq:C1}, which is central to the theory of this paper, converges to 0 as required. Our theory then implies that the permutation test using the regression residuals has asymptotically the same size as the permutation test on the true errors. Since, by construction, the errors are exchangeable given $X$, our theory suggests that the residual permutation test should be asymptotically valid. This is exactly what we see in the second column reporting the rejection rates. Indeed, the residual permutation test gets closer to the nominal level as $n$ grows. 

\renewcommand{\arraystretch}{1.2}
\begin{table}[h!]
\centering
\begin{tabular}{r r rr}
  \hline
\hline
$n$ & Rejection (\%) & Ratio in (C1) & Hoeffding's ratio~\eqref{eq:hoeff_leveraged}    \\ 
\hline
20    &      6.98     &   0.0860   &         0.606 \\
50     &       5.86     & 0.0348     &       0.613\\
100     &       5.24    &   0.0184     &       0.619\\
250     &       5.04     &  0.00760   &        0.623 \\
\hline
\end{tabular}
\caption{Rejection rates (\%) for the residual permutation test in the simulation of Example~\ref{example:hoeffding}. 
}
\label{tab:hoeff_perm}
\end{table}
\end{example}

\subsection{Residual sign randomization tests}
\label{sec:hs}
In this section, we consider the orthant symmetry invariance, $\Gall^\ss$, resulting in a residual sign test. This assumption may be useful in certain settings with heteroskedasticity; e.g., in a model with $\eps_i = f_i(X) U_i$, where $f_i$ is an arbitrary function and $U_i$ are i.i.d. symmetric random variables. Operationally, the residual sign test compares the observed value of the test statistic, $T_n$, with values of the statistic calculated at residuals with their signs randomly flipped. As pointed out by~\citet{diciccio2017robust}, this leads to a randomization-based analogue of the wild bootstrap procedure~\citep{wu1986jackknife, liu1988bootstrap, mammen1993bootstrap}.
 
Before we present our main result, we need to make some definitions. First, let $h_{ii}$ be the $i$-th diagonal element of the ``hat matrix'' $\Pn$, and define $\lev = \max_{i\in[n]} h_{ii} / (\sum_{j=1}^n h_{jj}/n)$. Also, let $\sigma_n^* = \max_{i\in[n]} \mE(\eps_i^2) / \min_{j\in[n]} \mE(\eps_j^2)$ and $\psi_n^* =  \max_{i\in[n]} \psi_i / \min_{i'\in[n]} \psi_{i'}$, where $\psi_i = (e_j' q_i)^2$ with $q_i= (\XnT X)^{-1} \XnT \ones_i$.  All these terms are leverage quantities over different aspects of the model. Concretely, $\lev$ is a summary of the overall design's leverage~\citep{huber1973robust}. See also~\citet{lei2021regression} for the large-$p$ regime. The term $\psi_n^*$ describes how leveraged the data are with respect to the particular covariate being tested; e.g., if $x_i$ are uncorrelated and we are testing one coefficient, then $\psi_n^* = \max_{i\in[n]} x_{ik}^2 / \min_{i'\in[n]} x^2_{i'k}$. Lastly, $\sigma_n^*$ is the condition number of the $\eps$'s covariance matrix.

\begin{theorem}\label{thm:s}
\TheoremS
\end{theorem}

Analogous to the results of the previous section, Theorem~\ref{thm:s} also has a dual interpretation. First, if $\eps$ are symmetric conditional on $X$, the theorem shows that the  residual sign test is asymptotically valid as long as the design is not too highly-leveraged according to~\eqref{eq:C1_sign}.  We note that this condition does not exclude high-dimensional settings where $p\to\infty$. If, in addition, Assumption~\eqref{A:gammas} holds and $\min\{\sigma_n^*, \psi_n^*\} = O(1)$, then the Type I error of of the residual sign test
can be bounded using Theorem~\ref{thm:main} as follows:
\begin{equation}
\mE(\phi_n) \le \alpha + O\big( (\lev p/n)^{\gamma/(2+\gamma)} \big).\nonumber
\end{equation}

We see that the rate slows down whenever $p$ grows with $n$ or the design leverage becomes excessive. Without high leverage ---e.g., when $X$ and $h_{ii}$  are uniformly bounded--- the condition in Condition~\eqref{eq:C1_sign} of Theorem~\ref{thm:s} reduces to $p/n=o(1)$, the exact same condition as in Theorem~\ref{thm:perm} under exchangeability. Roughly speaking, sign symmetry without excessive leverage is an invariance that behaves similarly to exchangeability.

Alternatively, if the errors are not symmetric but the limit hypothesis~\eqref{eq:limit_hypothesis_error} is true, the theorem states that the residual sign test based on the studentized test statistic has asymptotically the same size as the idealized sign test based on the true errors, assuming that condition~\eqref{eq:C1_sign} of the theorem is true. This condition is generally satisfied in regular population models;  e.g., if $x_i \sim N(0,1)$ i.i.d.~then $\bar\lambda_n = O_P(\log n)$~\citep{embrechts2013modelling}. If, in addition, the errors have bounded second moments, then the main condition reduces to $\log n\cdot p /n = o(1)$, which is satisfied because $p$ is typically fixed under~\eqref{eq:limit_hypothesis_error}. Then, the asymptotic validity of the residual sign test under the studentized statistic follows from the asymptotic validity of the studentized sign test using the true errors, which has already been established in prior work~\citep{janssen1999nonparametric}.

\begin{remark}
A similar duality of interpretation exists relative to bootstrap in the sign symmetry setting as in the exchangeability setting. For instance, the analysis of \citet[Theorem 5, pp. 1275]{wu1986jackknife} does not require invariance under~$\Gall^\ss$, but requires that
$$
\text{$\eps_i$ are uncorrelated and i.i.d.},~\max_{i\in[n]} \mE(\eps_i^2) = O(1),~\text{and}~\lev p = O(1).
$$
These assumptions in fact imply that Condition~\eqref{eq:C1_sign} 
holds true if $\min\{\sigma_n^*, \psi_n^* \} = o(n)$, which is a mild restriction. It holds, for instance, when $\mE(\eps_i^2)$ is bounded above and below, or when $X$ is uniformly bounded; c.f.~Condition (C4) in~\cite{wu1986jackknife}. Our analysis offers a complementary set of conditions for wild bootstrap's validity under symmetric errors.
\end{remark}

\section{Concluding remarks}
This paper developed a finite-sample bound for the size of approximate randomization tests. This allowed us to derive new conditions for the validity of such tests under invariance assumptions, while being able to leverage existing results of validity in settings where invariance does not hold but the test statistics converge asymptotically via studentization.
One benefit of our approach, which we didn't exploit in this paper, is that our main results hold for arbitrary invariances, $\Gall$. 
This could be helpful in settings where inference is challenging, such as regression models with a complex clustered error structure. Another interesting future direction would be to understand what type of invariances are appropriate with dependent data ---graphs or time series--- and adapt our results to such settings.

\section*{Acknowledgement}
I would like to thank Peter Bickel, Ivan Canay, Peng Ding, Dean Eckles,
Max Farrell, Avi Feller, Chris Hansen, Guido Imbens, Ed Kao, James MacKinnon, Lihua Lei, Paul
Rosenbaum, and Azeem Shaikh for valuable feedback and discussions. 

\singlespacing
\bibliographystyle{plainnat}
\bibliography{paper-ref}

\appendix
\small

\newcommand{\thmA}{2.1}
\newcommand{\invA}{(1)} 
\newcommand{\cVar}{(C1)}
\newcommand{\AI}{(A1)}
\newcommand{\AIb}{(A1')}
\newcommand{\AII}{(A2)}
\newcommand{\AIII}{(A3)}
\newcommand{\AIIIb}{(A3')}

\section{Proof of Theorem~1}

\subsection{Notation}\label{appendix:notation}
$\Unif(S)$ represents the uniform random distribution on set $S$;  `$\myeq{d}$' denotes equality in distribution. 
For an integer $k$, let $[k] = \{1, \ldots, k\}$ and $[k]_0$ be the same set including 0.
For some invariant set $\Gall$, 
we defined
\begin{align}\label{eq:def_LD}
\Delta_n & = t_n(G\heps) - t_n(G\eps), ~\Lambda_n =t_n(G'\eps) - t_n(G''\eps),
\end{align}
where $G, G', G''\sim\Unif(\Gall)$ i.i.d.;
$\eps$ are the ``true'' random variables and $\heps$ are the proxy variables.
Define $\LamStud = \Lambda_n/\sigma_{\LamStud}$ for the standardized $\Lambda_n$, 
and let  $\FLn$ be its cdf.

In the randomization tests, we use $G_1, \ldots, G_m$, where  $G_r\myeq{d} \Unif(\Gall), r=1,\ldots, m$, i.i.d.
For the infeasible test, define
\begin{equation}\label{eq:def_Tne}
\Tne = \big\{ t_n(\eps) \big\} \cup \big\{ t_n(G_r \eps): r \in [m] \big\}.
\end{equation}
and for the feasible test define
\begin{equation}\label{eq:def_Tner}
\Tner = \big\{ T_n \big\} \cup \big\{ t_n(G_r \heps): r \in [m] \big\}.
\end{equation}
Recall that under the null hypothesis, $T_n=t_n(\eps)$ and so $\Tner$ under the null hypothesis is equal to
\begin{equation}\label{eq:def_Tner_H0}
\Tner \myeq{H_0} \big\{ t_n(\eps) \big\} \cup \big\{ t_n(G_r \heps): r \in [m] \big\}.
\end{equation}
We see that under the null hypothesis, $t_n(\eps)$ is always included in the sets 
and so it will be convenient to use the convention $G_0=1$, the identity transformation.
Under this convention, let
\begin{align}
\LamMin & = \min\big\{  |t_n(G_r\eps) - t_n(G_{r'}\eps) | 
: r, r'\in[m]_0, r\neq r'\big\} ~\text{and}~,\nn\\
\DeltaMax & = \max_{r \in [m]} | t_n(G_r\heps) - t_n(G_r\eps) |.\nn
\end{align}
That is, $\LamMin$ is the minimum spacing in $\Tne$, and $\DeltaMax$ is the maximum  error in $\Tner$ under the null hypothesis.

\subsection{Main proof}
%
We first show that the infeasible (idealized) randomization test using the true $\eps$ is exact. 
Then, we prove Theorem~\thmA~by showing that  the feasible test is asymptotically similar to the infeasible one.
Our proof follows three steps:
\begin{enumerate}
\item Show finite-sample validity of infeasible test under the invariance hypothesis.
\item Prove a robustness property of the infeasible test.
\item For the feasible test analyze (a) Properties of $\DeltaMax$ and $\LamMin$; 
and (b) Its asymptotic validity.
\end{enumerate}

\noindent\underline{Step 1. Validity of infeasible randomization test.}
Suppose that $\Gall$ is arranged in some fixed arbitrary order, and define
 $$
 \mathbb{W}_m = \bigg\{ w \in \{0, 1\}^{|\Gall|}: \sum_{\gg} w_{\gg}=m+1, w_1=1\bigg\}.
 $$
 Let $W\in\mathbb{W}_m$ denote the binary vector `indexing' the transformations sampled in $\Tne$ and $\Tner$.
 It is easy to see that  $W \myeq{d} \Unif(\mathbb{W}_m)$, and that $W$ is independent of $\eps$. 
 
 We can now rewrite our sets using $W$:
 \begin{equation}\label{eq:def_Tne2}
\Tne =   \big\{ t_n(\gg \eps): \gg\in\Gall, W_{\gg}=1 \big\} := \TTne(\eps, W).
\end{equation}
This equivalence shows that the random part of $\Tne$ is uniquely determined by $\eps$ and $W$, which are mutually independent.
The critical value function in~(3) of the main text can be written as:
\begin{align}\label{eq:def_critical2}
c_{n,\alpha}(\Tne)  =  \inf\bigg\{ z \in\Real: \frac{1}{m+1} \sum_{\gg\in\Gall} W_{\gg}  \Indb{t_n(\gg\eps) \le z} \ge 1-\alpha\bigg\}
:= c_{\alpha}(\eps, W).
\end{align}
As before, this equivalence shows that the random part of the critical value depends only  on $\eps$ and $W$. 
The one-sided (infeasible) test can similarly be re-defined as
\begin{equation}\label{eq:def_infeasible2}
\phione^* = \Indb{T_n > c_{n,\alpha}(\Tne)} = \Ind\{T_n > c_{\alpha}(\eps, W)\}.
\end{equation}
\begin{lemma}\label{lemma:invar_properties}
For a fixed $\gg\in\Gall$, let  $\pi_{\gg}\in\mathsf{S}_{|\Gall|}$ be a permutation of $\Gall$ that maps element $\gg'\in\Gall$ to $\gg'\gg^{-1}$.
Then,
\begin{align}
\TTne(\gg\eps, W) & = \TTne(\eps, \pi_g W)~\nn\\
c_{\alpha}(\gg\eps, W)  & = c_{\alpha}(\eps, \pi_{\gg} W).
\end{align}
\end{lemma}
\begin{proof}
By definition, 
\begin{align}
\TTne(\eps, \pi_g W) = \{t_n(\gg' \eps) : g'\in\Gall, W_{\gg'\gg^{-1}}=1\} \myeq{(i)} \{t_n(\gg'' \gg \eps) : g''\in\Gall, W_{\gg''}=1\}
= \TTne(\gg\eps, W).\nn
\end{align}
Here, (i) follows from the group structure of $\Gall$. Similarly,
\begin{align}
c_{\alpha}(\gg\eps, W) & =  \inf\bigg\{ z \in\Real: \frac{1}{m+1}\sum_{\gg'\in\Gall}  W_{\gg'} \Indb{t_n(\gg'\gg\eps) \le z} \ge 1-\alpha\bigg\}\nn\\
& \myeq{(i)} \inf\bigg\{ z \in\Real: \frac{1}{m+1}\sum_{\gg''\in\Gall}  W_{\gg'' \gg^{-1}} \Indb{t_n(\gg''\eps) \le z} \ge 1-\alpha\bigg\} =c_{\alpha}(\eps, \pi_{\gg} W).\nn
\end{align}
As before, (i) follows from the group structure of $\Gall$. 
\end{proof}
%

\begin{lemma}\label{lemma:infeasible_valid}
For the infeasible test in~\eqref{eq:def_infeasible2}, the null hypothesis 
implies 
$$
\mE(\phione^*)  \le \alpha,
$$
for all $n>0$ if the invariance property on $\eps$ holds true.
\end{lemma}
\begin{proof}
 Consider the function  $d(\eps, W) =  \Ind\{t_n(\eps) > c_{\alpha}(\eps, W)\}$.
From this we obtain
\begin{align}\label{lemma:infeasible_valid-1}
\sum_{\gg\in\Gall} W_{\gg} d(\gg\eps, \pi_{\gg^{-1}} W)   & = \sum_{\gg} 
W_{\gg}  \Ind\{t_n(\gg\eps) > c_{\alpha}(\gg\eps,  \pi_{\gg^{-1}} W)\}  \nn\\
 & \myeq{(i)} \sum_{\gg} W_{\gg} \Ind\{ t_n(\gg\eps) > c_{\alpha}(\eps, W) \} =
     \sum_{\gg} W_{\gg} - \sum_{\gg} W_{\gg} \Ind\{ t_n(\gg\eps) \le c_{\alpha}(\eps, W) \}   \myle{(ii)} (m+1) \alpha.
\end{align}
Here, (i) follows from~Lemma~\ref{lemma:invar_properties} and $\pi_{\gg}  \pi_{\gg^{-1}} = 1$;
 and (ii) follows from~\eqref{eq:def_critical2} and from $\sum_{\gg} W_{\gg} = m+1$.

Next, fix  some $\gg\in\Gall$ for which $W_g=1$, and let $\eps' = g \eps$ and $W'$ be constructed from $W$ such that 
$W'_{g'}= 1$ if and only if $W_{g' g^{-1}}=1$.  As such, there is a one-to-one mapping between $W'$ and $W$ since 
$W' = \pi_{g^{-1}} W$. 
We claim that
\begin{align}\label{lemma:infeasible_valid-2}
\Ind\{t_n(\eps) > c_{\alpha}(\eps, W)\} \myeq{d} \Ind\{t_n(\eps') > c_{\alpha}(\eps', W')\}  = 
\Ind\{t_n(\eps') > c_{\alpha}(\eps, W)\} .
\end{align}
To prove the first part of the claim, note that $(\eps', W') \myeq{d} (\eps, W) \mid W_g=1$. 
This  follows from the invariance property of $\eps$, the definition of $W$, and the independence between $\eps, W$.
The second part of the claim follows immediately from 
$ c_\alpha(\eps', W') = c_\alpha(g\eps, \pi_{g^{-1}} W) = c_\alpha(\eps, W)$, where 
in the last step we applied Lemma~\ref{lemma:invar_properties}.

It now follows that
\begin{align}\label{lemma:infeasible_valid-3}
W_{\gg} d(\gg \eps, \pi_{\gg^{-1}} W) 
& = W_{\gg} \Ind\{t_n(\gg\eps) >  c_\alpha(\gg \eps, \pi_{\gg^{-1}} W) \} \nn\\
& \myeq{(i)} W_{\gg} \Ind\{t_n(\gg\eps) > c_\alpha(\eps, W) \} \nn\\
&\myeq{(ii),d} W_{\gg} \Ind\{t_n(\eps) > c_\alpha(\eps, W) \} \nn\\
& = W_{\gg}  d(\eps, W).
\end{align}
Here, (i) follows from Lemma~\ref{lemma:invar_properties} and (ii) from~\eqref{lemma:infeasible_valid-2}
Finally, 
\begin{align}
 \mE(\phione^*) = P\{T_n > c_{n,\alpha}(\Tne)\} 
 & \myeq{(i)} \mE(\Ind\{t_n(\eps) > c_{\alpha}(\eps, W)\})  \nn\\
 & = \mE[d(\eps, W)] \myeq{(ii)} \frac{1}{m+1} \sum_{\gg\in\Gall} \mE[ W_{\gg} d(\eps, W)] \nn\\
 & \myeq{(iii)}  \frac{1}{m+1} \sum_{\gg\in\Gall} \mE[ W_{\gg} d(\gg\eps, \pi_{\gg^{-1}} W)]   \myle{(iv)} \alpha.\nn
\end{align}
Here, (i) follows from $T_n=t_n(\eps)$ under the null hypothesis; 
(ii) follows from $\sum_{\gg} W_{\gg} = m+1$ almost surely; (iii) follows from ~\eqref{lemma:infeasible_valid-3}; and (iv) follows from~\eqref{lemma:infeasible_valid-1}.
\end{proof}

\vspace{5px}
\noindent\underline{Step 2. Robustness of the idealized randomization test.}
Here, we show that the idealized test is robust to certain ``corruptions'' of the randomization distribution. 
Later, we will show that this robustness  can ``withstand'' the level of corruption 
introduced in the feasible test by using the proxy variables instead of the true variables.

Define $\xi: \Real^{m+1} \to  \Real$ as $\xi(\{t_1, \ldots, t_{m+1}\}) = \min_{j=1, \ldots, m} \{ t_{(j+1)} - t_{(j)} \}$,
where $t_{(j)}$ is  the $j$-th ordered value without repetitions. As such, $\xi$ calculates the minimum spacing in a set of values. 
Assume $M$ unique values, i.e., $t_{(1)}$ is the smallest value and $t_{(M)}$ is the largest. With this notation, we can write
\begin{equation}\label{eq:def_Lnstar}
\LamMin = \xi(\Tne)
\end{equation}
as the minimum spacing in the randomization values of the idealized test. We set $\LamMin=0$ when all randomization values are equal.
The following result quantifies how a corruption at the order of $\LamMin$ affects the rejection rate of the idealized test.

\begin{lemma}\label{lemma:xi}
Under the null hypothesis,
$$
P\left\{T_n  > c_{n, \alpha}(\Tne)  - \LamMin \right\} \le \alpha + O(1/m) + O(m \gon) + O(m^2/ |\Gall |).
$$
\end{lemma}
\begin{proof}
Since $c_{\alpha}$ is an order statistic in $\Tne(\eps)$ 
and $\LamMin$ is the minimum spacing in $\Tne(\eps)$, the following relationship holds as an identity:
\begin{equation}\label{lemma:xi-1}
W_{\gg} \Ind\{ t_n(\gg\eps) > c_{\alpha}(\eps, W) -  \LamMin \}
\le W_{\gg} \Ind\{ t_n(\gg\eps) > c_{\alpha}(\eps, W) \} + W_{\gg} \Ind\{t_n(\gg\eps) = c_{\alpha}(\eps, W)\}.
\end{equation}
This leads to
\begin{align}\label{lemma:xi-2}
\sum_{\gg\in\Gall} W_{\gg} \Ind\{ t_n(\gg\eps) > c_{\alpha}(\eps, W) -  \LamMin\}
&\myle{(i)}  \sum_{\gg} W_{\gg} \Ind\{ t_n(\gg\eps) > c_{\alpha}(\eps, W)\} +  \sum_{\gg} W_{\gg} \Ind\{t_n(\gg \eps)  =  c_{\alpha}(\eps, W)\}\nn\\
&\myle{(ii)} (m+1) \alpha + R_{n,\alpha},
\end{align}
where $R$ is the multiplicity of randomization values exactly at the $\alpha$-critical value.
Here, (i) follows from~\eqref{lemma:xi-1}, and (ii) follows from~\eqref{eq:def_critical2}. 
Consider the  decision function 
$$
d'(\eps, W) =  \Ind\{t_n(\eps) > c_{\alpha}(\eps, W) -  \LamMin\} =  \Ind\{t_n(\eps) > c_{\alpha}(\eps, W) -  \xi(\TTne(\eps, W))\}.
$$
We then obtain
\begin{align}\label{lemma:xi-3}
\sum_{\gg\in\Gall} W_{\gg} d'(\gg\eps, \pi_{\gg^{-1}} W)   & = \sum_{\gg} 
W_{\gg}  \Ind\{t_n(\gg\eps) > c_{\alpha}(\gg\eps,  \pi_{\gg^{-1}} W) - \xi(\TTne(\gg\eps, \pi_{\gg}^{-1}W))\}  \nn\\
 & \myle{(i)}  \sum_{\gg} W_{\gg} \Ind\{ t_n(\gg\eps) > c_{\alpha}(\eps, W) - \xi(\TTne(\eps, W))\} \nn\\
& \myle{(ii)}  (m+1) \alpha + R_{n,\alpha}.
\end{align}
In the above derivation, (i) follows from~Lemma~\ref{lemma:invar_properties} and $\pi_{\gg}  \pi_{\gg^{-1}} = 1$, the identity permutation;
 and (ii) follows from~\eqref{lemma:xi-2}.
 Following the same argument as in the proof of~\eqref{lemma:infeasible_valid-3} of Lemma~\ref{lemma:infeasible_valid} we can show that
 \begin{align}\label{lemma:xi-4}
\mE[ W_{\gg} d'(\gg\eps, \pi_{\gg^{-1}} W) ]  = \mE[ W_{\gg} d'(\eps, W)].
\end{align}
Under the null hypothesis, 
\begin{align}
P\{T_n >  c_{n,\alpha}(\Tne) - \LamMin\} 
 & \myeq{(i)} \mE(\Ind\{t_n(\eps) > c_{\alpha}(\eps, W) -  \LamMin\}) 
 = \mE[d'(\eps, W)] \nn\\
  & \myeq{(ii)} \frac{1}{m+1} \sum_{\gg\in\Gall} \mE[ W_{\gg} d'(\eps, W)]\nn\\
  &  \myeq{(iii)} \frac{1}{m+1} \sum_{\gg}  \mE[W_{\gg} d'(\gg\eps,  \pi_{\gg^{-1}} W)] 
 \myle{(iv)} \alpha + \mE(R_{n, \alpha})/ (m+1).
\end{align}
Here, (i) follows from $T_n=t_n(\eps)$ under the null hypothesis; 
(ii) follows from $\sum_{\gg} W_{\gg}=m+1$; (iii) follows from~\eqref{lemma:xi-4} and (iv) follows from~\eqref{lemma:xi-3}. 

Next, we derive a bound on the multiplicity term. Let 
$G_0=1$ be the identity transformation. 
Let $E_n=1$ denote the event that $G_r \neq G_{r'}\neq 1$, for all~$r,r'=0, \ldots, m, r\neq r'$, and let $E_n=0$ otherwise.
Since $G_r$ for $r>0$s are i.i.d. uniform from $\Gall$,
\begin{align}\label{lemma:xi-5}
P(E_n=1) = \frac{1}{1}  \cdot  \frac{|\Gall |  -1}{|\Gall | } \cdots \frac{|\Gall | -m}{|\Gall|} 
\ge [1-m/ |\Gall|]^{m} \myge{(i)} 1-m^2 /|\Gall|.
\end{align}
Here, (i) follows for some sufficiently large $n$ since $m = O(1)$ and $|\Gall| \to\infty$. 
Define $c_n = c_{n,\alpha}(\Tne)$ for brevity. 
Then,
\begin{align}
R_{n,\alpha} & =  \sum_{r=0}^m  \Indb{t_n(G_r\eps) = c_n}  
=  \sum_{r=0}^m  \Indb{t_n(G_r\eps) = c_n}  (\Ind\{E_n=1\} + \Ind\{E_n=0\})  \nn\\
& \le \Ind\{E_n=1\}  \sum_{r=0}^m  \Indb{t_n(G_r\eps) = c_n} + (m+1) \Ind\{E_n=0\}  \nn\\
& \le  1+\sum_{r\neq r', G_r\neq G_{r'}}  \Indb{ t_n(G_r\eps) = t_n(G_{r'}\eps)} +   (m+1) \Ind\{E_n=0\} \nn\\
\mE(R_{n,\alpha}) &  \myle{(i)} 1+ \sum_{r\neq r', G_r\neq G_{r'}} P\{t_n(G_r\eps)= t_n(G_{r'} \eps) \} + (m+1)m^2 /|\Gall |, \nn\\
& \myle{(ii)} 1+ m^2 \gon +  (m+1)m^2 /|\Gall | \nn
\end{align}
and so $[1/(m+1)] \mE(R_{n,\alpha}) \le O(1/m) + O(m \gon) + O(m^2/ |\Gall |)$. 
In the above derivations, result (i) follows from~\eqref{lemma:xi-5} and (ii) follows by definition of $\gon$.
\end{proof}

\noindent\underline{Step 3. Feasible  randomization test.}

Here, we turn our focus to the feasible test:
\begin{align}\label{actual_phi}
\phi_n &= \Ind\{T_n > c_{n, \alpha}(\Tner)\}.
\end{align}
Our proof strategy relies on showing that the feasible test satisfies a bound of the form
$$
\mE(\phi_n) \le \alpha + o(1) + P(\DeltaMax  > \LamMin),
$$
as defined in~Section~\ref{appendix:notation}.
We will show that Condition~\cVar~ is sufficient to guarantee that $P(\DeltaMax > \LamMin) = o(1)$. The following property of the critical value will be useful.
\begin{lemma}\label{lemma:c_prop}
The critical value function satisfies
\begin{align}
c_{n,\alpha}(\{t_1 - \delta, \ldots, t_{m+1} - \delta\})  = c_{n,\alpha}(\{t_1, \ldots, t_{m+1}\}) - \delta,~\text{for any}~\delta > 0.\nonumber
\end{align}
\end{lemma}
\begin{proof}
By the infimum definition, the critical value is equal to $t_{(j)}$ for some unique $j=1, \ldots, M$. By linearity of ranks, 
$c_{n, \alpha}(\{t_1-\delta, \ldots, t_{m+1}-\delta\})$ is equal to $t_{(j)}-\delta$.
\end{proof}

\vspace{5px}
\noindent\underline{Step 3a. Properties of $\DeltaMax$ and $\LamMin$.}~The following lemma connects the minimum spacing $\LamMin$ in the infeasible randomization test 
to the spacings variable, $\Lambda_n$,  which is easier to analyze.
\begin{lemma}\label{lemma:Lnstar}
Let $\sigma_{\Lambda_n}^2 = \var(\Lambda_n)$ where  $\Lambda_n = t_n(G\eps) - t_n(G'\eps)$, $G, G'\sim\Unif(\Gall)$, 
and define $\LamStud= \Lambda_n/\sigma_{\Lambda_n}$ as its standardization. Then, for any $\epsilon>0$, 
\begin{align}\label{eq:Xin_1}
P( \LamMin /\sigma_{\Lambda_n} < \epsilon) \le O(m^2) [\FLn(\epsilon) - \FLn(-\epsilon)].
\end{align}
\end{lemma}
\begin{proof}
The event $\{\LamMin /\sigma_{\Lambda_n} < \epsilon\}$ implies that for some randomizations $G_r, G_{r'}$, $r\neq r'$, 
it happened that $|t_n(G_r \eps) - t_n(G_{r'}\eps) |/\sigma_{\Lambda_n} < \epsilon$. Therefore, we can ``envelope" this event with the following expression
$$
\{ \LamMin / \sigma_{\Lambda_n} < \epsilon \}  \subseteq
\bigcup_{r, r'\in\{0, 1, \ldots, m\}, r\neq r'} \{  | t_n(G_r\eps) - t_n(G_{r'} \eps) | / \sigma_{\Lambda_n} < \epsilon \},
$$
where we assume $G_0=1$, as before. It follows that
\begin{align}\label{eq2708}
P( \LamMin / \sigma_{\Lambda_n} < \epsilon) & \le \sum_{ r\neq r'} P\{ | t_n(G_r\eps) - t_n(G_{r'} \eps)|  / \sigma_{\Lambda_n} < \epsilon\}  \nn\\
 &\myle{(i)}  \sum_{ r\neq r'} P\{ | \Lambda_n |  / \sigma_{\Lambda_n} < \epsilon\} 
= O(m^2) P\{ | \Lambda_n |  / \sigma_{\Lambda_n} < \epsilon\}  \nn\\
 & = O(m^2) [\FLn(\epsilon) - \FLn(-\epsilon)].
\end{align}
Here, (i) follows from independence of $G_{r}, G_{r'}$ when $r\neq r'$.
\end{proof}

The following lemma establishes a useful result on the distribution of $|\Lambda_n|$ near zero.
\begin{lemma}\label{lemma:F_Lnstar}
For variable $\LamStud$, under Assumption~(A2), it holds that
\begin{align}
\lim_{\epsilon\to0^+} [\FLn(\epsilon) - \FLn(-\epsilon)] \le 3(\gon + 1/|\Gall|).\nn
\end{align}
\end{lemma}
\begin{proof}
First, we obtain
\begin{align}\label{lemma:F_Lnstar-1}
P(\LamStud=0) = P(\Lambda_n=0) & = P\{t_n(G\eps)=t_n(G'\eps)\} \nn\\
& \myeq{(i)} (1/|\Gall|^2) \sum_{\gg,\gg'\in\Gall} P\{ t_n(\gg\eps)=t_n(\gg'\eps) \} \nn\\
&  =  (1/|\Gall|^2) \sum_{\gg=\gg'} 1 +  (1/|\Gall|^2) \sum_{\gg\neq \gg'}  P\{t_n(\gg\eps)=t_n(\gg'\eps)\} \nn\\
& \myle{(ii)}(1/|\Gall|) +  (1/|\Gall|^2) \sum_{\gg\neq \gg'}  \gon  \le \gon + 1/|\Gall|.
 \end{align}
Here, (i) follows from the i.i.d.~distribution of $G, G'$; (ii) follows by definition of $\gon$ in Assumption~(A2).
Next, for any $t\ge 0$ define $\mathbb{S}_t(\eps) = \{ (\gg,\gg')\in\Gall \times \Gall :
 |t_n(\gg\eps) - t_n(\gg'\eps) | / \sigma_{\Lambda_n} \le t \}$ and
$$
t^*(\eps) = 
\begin{cases}
~~~~~~\quad\quad\quad\quad\quad\quad\quad\quad\quad\quad\quad\quad0 ,  &    \text{if}~ \Gall \times \Gall \setminus \mathbb{S}_0(\eps) = \emptyset\\
 \min\big\{  |t_n(\gg\eps) - t_n(\gg'\eps) | / \sigma_{\Lambda_n}: (\gg, \gg') \in \Gall \times \Gall \setminus \mathbb{S}_0(\eps) \big\}  > 0, & \text{otherwise}.
\end{cases}
$$
Intuitively, $t^*(\eps)= 0$ whenever all test statistic values are equal across all pairs from $\Gall$ given $\eps$;  
otherwise, it takes the minimum nonzero discrepancy.
This definition implies the following identity ---for any $G, G'\in\Gall$, $\epsilon>0, \eps\in\Real^n$---
\begin{align}\label{lemma:F_Lnstar-2}
 \Indb{ |t_n(G\eps) - t_n(G'\eps)| / \sigma_{\Lambda_n} \le \epsilon } \Ind\{ t^*(\eps)\neq0\}
 \le \Ind\{t^*(\eps) \le \epsilon\} + \Ind\{t_n(G\eps) = t_n(G'\eps) \}.
 \end{align}
Moreover, 
 \begin{align}\label{lemma:F_Lnstar-3}
\lim_{\epsilon\to 0^+} P\{t^*(\eps) \le \epsilon\} & \myeq{(i)} P\{t^*(\eps) = 0\} \nn\\
& = \mE\big[ \Ind\{t_n(\gg\eps)=t_n(\gg'\eps) : \forall~\gg,\gg'\in\Gall \} \big] \nn\\
& \myle{(ii)}   \mE\big[  \sum_{\gg,\gg'\in\Gall} \Ind\{t_n(\gg\eps)=t_n(\gg'\eps) \} / |\Gall|^2 \big]\nn\\
&   = P(\Lambda_n=0) \myle{(iii)} \gon + 1/|\Gall|.
 \end{align}
 Here, (i) follows from right continuity of cdfs; (ii) holds as an identity since $\Ind\{t_n(\gg\eps)=t_n(\gg'\eps) : \forall~\gg,\gg'\in\Gall\}=1$
 implies $\sum_{\gg,\gg'\in\Gall} \Ind\{t_n(\gg\eps)=t_n(\gg'\eps) \} /|\Gall|^2 = 1$; and (iii) follows from~\eqref{lemma:F_Lnstar-1}. Thus, for any $\epsilon>0$ we obtain
\footnotesize{
\begin{align}
\Indb{ \frac{|t_n(G\eps) -  t_n(G'\eps)| }{ \sigma_{\Lambda_n} } \le\epsilon} & =
 \Indb{ \frac{|t_n(G\eps)  - t_n(G'\eps)| }{\sigma_{\Lambda_n}} \le\epsilon } \Ind\{ t^*(\eps)=0\} + 
 \Indb{ \frac{ |t_n(G\eps) - t_n(G'\eps)| }{ \sigma_{\Lambda_n}}  \le\epsilon} \Ind\{ t^*(\eps)\neq0\} \nn\\
& \myle{(i)} \Indb{ t_n(G\eps) = t_n(G'\eps) }  +  \Indb{\frac{ |t_n(G\eps) - t_n(G'\eps)|}{ \sigma_{\Lambda_n}}  \le\epsilon } \Ind\{ t^*(\eps)\neq0\} \nn\\
&   \myle{(ii)} 2 \Indb{ t_n(G\eps) = t_n(G'\eps) }  +  \Ind\{t^*(\eps) \le \epsilon\} \nn\\
P\big\{ \frac{|t_n(G\eps) - t_n(G'\eps)|}{ \sigma_{\Lambda_n}}  \le\epsilon \big\} & \le 2 P\{t_n(G\eps) = t_n(G'\eps) \}  +  P\{ t^*(\eps) \le \epsilon\} \nn\\
& = 2 P\{ \Lambda_n=0 \}  +  P\{ t^*(\eps) \le \epsilon\} \nn\\
 \FLn(\epsilon) - \FLn(-\epsilon)& \myle{(iii)} 2(\gon + 1/|\Gall|) +  P\{ t^*(\eps) \le \epsilon\}.\nn
 \end{align}
 }
 \normalsize
 Here, (i) follows from definition of $t^*(\eps)$ which implies that $t_n(\gg\eps)=t_n(\gg'\eps)$ for all $\gg, \gg'\in\Gall$ whenever $t^*(\eps)=0$;
 (ii) follows from~\eqref{lemma:F_Lnstar-2}; (iii) follows from~\eqref{lemma:F_Lnstar-1}. 
 The final result follows from~\eqref{lemma:F_Lnstar-3} by taking $\epsilon\to 0^+$.
\end{proof}

We now derive a property for $\DeltaMax$ based on $\Delta_n=|t_n(G\heps)-t_n(G\eps)|$, the approximate error 
between the feasible and the infeasible test. 

\begin{lemma}\label{lemma:Dn}
Let $\mu_n = \mE(\DeltaMax / \sigma_{\Lambda_n})$ and $\zeta_n^2 = \mE(\Delta_n^2) ~/~ \mE(\Lambda_n^2)$.
Then, for any $\epsilon>0$, 
\begin{align}\label{eq:Dn}
P(\DeltaMax / \sigma_{\Lambda_n} - \mu_{n} > \epsilon)  \le m \zeta_n^2/\epsilon^2.
\end{align}
\end{lemma}
\begin{proof}
Let $\Delta_{n,r} = t_n(G_r\heps) - t_n(G_r\eps)$, where $G_r$ were sampled i.i.d in the randomization test, 
$r=1, \ldots, m$.  Since $G_r$ are independently drawn,
 $\mE(\Delta_{n,r}^2) = \mE(\Delta_n^2)$, and so
\begin{equation}\label{lemma:Dn-1}
\mE(\DeltaMax^2) \myle{(i)} \sum_{r=1}^m \mE(\Delta_{n,r}^2) = m\mE(\Delta_n^2).
\end{equation}
Here, (i) follows from $\DeltaMax^2 \le \sum_r |\Delta_{n,r}|^2$ since $\DeltaMax$ is one of $|\Delta_{n,r}|$.
Thus,
$$
P(\DeltaMax / \sigma_{\Lambda_n}  - \mu_{n} > \epsilon) \le P( |\DeltaMax / \sigma_{\Lambda_n}  -\mu_{n}| > \epsilon) 
\myle{(i)} \mE(\DeltaMax^2 / \sigma_{\Lambda_n}^2)/ \epsilon^2 \myle{(ii)}  \frac{m \mE(\Delta_n^2)}{\mE(\Lambda_n^2)} (1/\epsilon^2) 
= m\zeta_n^2/\epsilon^2.
$$ 
Here, (i) follows from Chebyshev's inequality, and (ii) from~\eqref{lemma:Dn-1}. 
and $\mE(\Lambda_n^2) = \var(\Lambda_n)$ as $\mE(\Lambda_n)=0$.
\end{proof}

The following result shows that $\DeltaMax$ is dominated by $\LamMin$ under Condition~\cVar.
\begin{lemma}\label{lemma:DLnstar}
Suppose that Assumptions~\AI-\AIII~hold. 
Then, for any $\epsilon > \sqrt{m}\zeta_n$,
$$
P(\DeltaMax >  \LamMin) \le m \zeta_n^2 /\epsilon^2 + O(m^2) [\FLn(2\epsilon) - \FLn(-2\epsilon)].
$$
Under  Condition~\cVar, it follows that 
$
P(\DeltaMax >   \LamMin)  = o(1).
$
If, in addition, Assumption~\AIIIb~holds then
$$
P(\DeltaMax >   \LamMin) \le O(m^2)(\gon + 1/|\Gall |) + A(\gamma, m) O(\zeta_n^{2\gamma/(2+\gamma)}),
$$
where $A(\gamma, m) = 8(1+\gamma) m^{(4+\gamma) / (2+\gamma)}$.
\end{lemma}
\begin{proof}
For any $x>0$ let $Q(x) = \FLn(x) - \FLn(-x)$.  First, we obtain
\begin{align}\label{lemma:DLnstar-1}
 P(\DeltaMax >   \LamMin)
  & \myeq{(i)} P(\DeltaMax / \sigma_{\Lambda_n} - \mu_{n} >  \LamMin/\sigma_{\Lambda_n} -\mu_{n}) \nn\\
  & \myle{(ii)}  P(\DeltaMax / \sigma_{\Lambda_n} - \mu_{n} > \epsilon) + P( \LamMin/\sigma_{\Lambda_n} - \mu_{n} < \epsilon) \nn\\
   & \myle{(iii)}  m \zeta_n^2/\epsilon^2 +  O(m^2) Q(\mu_{n}+\epsilon).
\end{align}
Here, (i) takes $\mu_{n}, \sigma_{\Lambda_n}$ as defined in Lemma~\ref{lemma:Dn};
(ii) follows from indicator identity $\Ind\{d > \lambda\} \le \Ind\{d > \epsilon\} + \Ind\{\lambda < \epsilon\}$;
the first  part of (iii) follows from Lemma~\ref{lemma:Dn}, and the second part follows from 
Lemma~\ref{lemma:Lnstar}. 
%
By Jensen's inequality,
$\mE(\DeltaMax^2 / \sigma^2_{\Lambda_n}) \ge \mu_{n}^2$, and so from~\eqref{lemma:Dn-1}  we 
 get $m \mE(\Delta_n^2) / \sigma^2_{\Lambda_n} \ge \mu^2_{n}$, and $m \zeta_n^2 \ge \mu^2_{n}$. 
  Therefore, for any $\epsilon \ge  \sqrt{m} \zeta_n$ it also holds $\epsilon \ge \mu_n$.  Since $Q(x)$ is non-decreasing, this allows us to simplify the bound to:
\begin{align}\label{lemma:DLnstar-2}
 P(\DeltaMax >   \LamMin) \le  m \zeta_n^2 /\epsilon^2 + m^2 Q(2\epsilon),
\end{align}
for any $\epsilon > \sqrt{m}\zeta_n$.\end{proof}

Finally, under Condition~\cVar~ we have $\zeta_n = o(1)$.
Assumption~(A3) implies that $\lim_{\epsilon\to0+} Q(x) = 0$ 
uniformly for any decreasing sequence $\epsilon$. Thus, Assumptions~\AI-\AIII~imply that in Eq.~\eqref{lemma:DLnstar-2} we can decrease $\epsilon$ at an appropriate rate~(slower than $\zeta_n$)
such that the upper bound vanishes. From this we obtain $ P\{\DeltaMax >  \LamMin\} = o(1)$.

\vspace{5px}
\noindent\underline{Step 3b. Finishing the proof.}~
Now, we put the pieces together. Let $\Delta_{n,r} = t_n(G_r\heps) - t_n(G_r\eps)$, for $r=1, \ldots, m$ 
with $\Delta_{n,0}=0$. Suppose the invariance condition on $\eps$ holds true. Then, under the null hypothesis, we obtain
\begin{align}\label{eq:one-sided_bound}
E(\phi_n) = P\left\{T_n > c_{n,\alpha}(\Tner)\right\} & \myeq{(i)} 
P\left\{T_n  > c_{n,\alpha}(\{ t_n(G_r \eps) +\Delta_{n,r} :  r=0, \ldots, m\})\right\}
\nonumber\\
 & \le P\left\{T_n > c_{n,\alpha}(\{  t_n(G_r \eps)  - \max_r |\Delta_{n,r}| :  r=0, \ldots, m\}) \right\}
  \nonumber\\
   & \myeq{(ii)} P\left\{T_n  > c_{n,\alpha}(\Tne)  - \DeltaMax\right\}
  \nonumber\\
               & \myle{(iii)} P\left\{T_n   > c_{n,\alpha}(\Tne)  -  \LamMin \right\}
        +  o(1).\nn\\
   & \myle{(iv)}  E(\phi_n^*) + O(1/m) +  P(\DeltaMax >  \LamMin).
\end{align}
Here, (i) follows from~\eqref{eq:def_Tner_H0}; (ii) follows from Lemma~\ref{lemma:c_prop} and the definition of $\DeltaMax$; in (iii)  we used the indicator identity $\Ind\{t > c -d\} \le \Ind\{t > c-\lambda\} + \Ind\{d > \lambda\}$; (iv) follows from Lemma~\ref{lemma:xi} and Lemma~\ref{lemma:DLnstar} under Condition~(C1). We can follow an identical process to prove the result for the two sided test, concluding the proof of Theorem 1.


\section{Proof of Theorem~2}
\begin{proof}
Under Assumption~\AIIIb,  we can optimize~\eqref{lemma:DLnstar-2} with respect to $\epsilon$ to obtain the tightest bound possible.
Minimizing (without constraints) a series of the form $A \zeta_n^2/\epsilon^2 + B \epsilon^\gamma$, where $A, B, \gamma>0$ yields the solutions
\begin{align}\label{lemma:DLnstar-3}
\arg\min_{\epsilon>0} \{ A \zeta_n^2/\epsilon^2 + B \epsilon^\gamma \} & = O(\zeta_n^{\gamma/(2+\gamma)}),~\text{and, more specifically,} \nn\\
\min_{\epsilon>0} \{ A \zeta_n^2/\epsilon^2 + B \epsilon^\gamma \} & = [A^\gamma B^2]^{1/(2+\gamma)} (1+\gamma/2) (2/\gamma)^{\gamma/(2+\gamma)}  \zeta_n^{2\gamma/(2+\gamma)}
\le [A^\gamma B^2]^{1/(2+\gamma)} (1+\gamma) \zeta_n^{2\gamma/(2+\gamma)}.
\end{align}
Since $\zeta_n=o(1)$ under Condition~\cVar~, it follows that $\zeta_n^{\gamma/(2+\gamma)} > \sqrt{m}\zeta_n$ for any sufficiently large $n>0$.
Then, the above solutions can be attained under the constrained formulation as well:
\begin{align}\label{lemma:DLnstar-4}
\min_{\epsilon>\sqrt{m} \zeta_n} \{ A \zeta_n^2/\epsilon^2 + B \epsilon^\gamma \} & \le [A^\gamma B^2]^{1/(2+\gamma)} (1+\gamma) \zeta_n^{2\gamma/(2+\gamma)}.
\end{align}
This implies that
\begin{align}\label{lemma:DLnstar-5}
 P(\DeltaMax >   \LamMin) & \le \inf_{\epsilon >\sqrt{m} \zeta_n} \big\{ m \zeta_n^2 /\epsilon^2 + m^2 Q(2\epsilon) \big\} \nn\\
 &  \myle{(i)} m^2 O(\gon +1/|\Gall|) + \inf_{\epsilon >\sqrt{m} \zeta_n} \big\{ m \zeta_n^2 /\epsilon^2 + m^2 2^\gamma\epsilon^\gamma \big\} \nn\\
 &  \myeq{(ii)} m^2 O(\gon +1/|\Gall|) + A(\gamma, m) \zeta_n^{2\gamma/(2+\gamma)}.
\end{align}
In the derivations above, (i) follows from~Assumption~\AIIIb, and (ii) from~\eqref{lemma:DLnstar-4} by setting $A=2m$ and $B=m^2 2^\gamma$. 
This implies a factor $A(\gamma, m) =  8(1+\gamma) m^{(4+\gamma) / (2+\gamma)}  = O(m^2)$. 

We can then plug in this bound in the proof of Theorem 1 to finish the proof of Theorem 2.
\end{proof}

\subsection{Numerical study: $\gamma$ parameter}
In this simulation, we numerically estimate $\gamma$ in Assumption~\AIIIb\ to understand its role in our theoretical results. Our working model is linear regression $y=X\beta + \eps$, and use the least squares estimator as the test statistic.
We set $p=5$, $n=5,000$ and estimate $\gamma$ over $50,000$ samples of $\eps$ conditional on $X$, and over 50 replications. The numerical results below show that $\gamma$ gets smaller for heavier-tailed data, especially responding more to the tails of the error distribution.
\renewcommand{\arraystretch}{1.5}
\begin{table}[h]
\centering
\begin{tabular}{ll | l}
$X$ & $\eps$ & $\gamma$ \\
\hline
N(0,1)        & N(0,1)  & 1.015 \\
               & Weibull(0.5) & 1.027 \\
               & Cauchy  & 0.337 \\
               \hline\hline
Weibull(0.5)        & N(0,1) & 1.019 \\
               & Weibull & 1.016 \\
               & Cauchy  & 0.344\\
               \hline
\end{tabular}
\end{table}

\section{Proof of Theorem~3}
\label{appendix:power}
\begin{proof}
Let $\Xin$ denote the range of  $\Tne = \{t_n(\eps)\} \cup \{t_n(G_r \eps)\}$, i.e., the difference between the maximum and minimum of the randomization values of the idealized test. Then,
for any $\alpha' \in [0, 1], n > 0$, it holds
\begin{equation}\label{thm:power-1}
 \Indb{ t_n(\eps)  \ge  c_{n,\alpha'}(\Tne) -\Xin } = 1.
\end{equation}
 By definition, $\Xin \ge  \LamMin$ with equality only 
in the extreme case where $\Xin = \LamMin=0$, that is,  all randomization values in $\Tne$ are identical. 
Then, 
 \begin{align}\label{thm:power-2}
 P(\DeltaMax \ge \Xin) \le P(\DeltaMax > \LamMin) + P(\LamMin =\Xin) 
& =  P(\DeltaMax > \LamMin) + P(\LamMin =0)  
 \nn\\
&  \myeq{(i)} 
  P(\DeltaMax > \LamMin) + O(m^2) (\gon + 1/|\Gall|) 
 \nn\\
 &  \myeq{(ii)} 
 O(m^2) (\gon + 1/|\Gall|) + A(\gamma, m)\cdot O(\zeta_n^{2\gamma/(2+\gamma)}) 
 \end{align}
 where (i) follows 
\eqref{eq2708} and Lemma~\ref{lemma:F_Lnstar}, 
and (ii) follows from Lemma~\ref{lemma:DLnstar}.
Moreover, for any $\tau>0$, the event $\Xin > \tau/2$ implies that 
there exists a pair in $\Tne$ for which the difference is  bigger than $\tau/2$, that is,
$$
\big\{ \Xin > \tau/2  \big\}  \subseteq
\bigcup_{r\neq r'} \big\{  | t_n(G_r\eps) - t_n(G_{r'}\eps) |  > \tau/2  \big\}
~ \bigcup_{r} \big\{  | t_n(G_r\eps) - t_n(\eps) |  > \tau/2  \big\}.
$$

Let's assume for now that the invariance property on $\eps$ holds. After laying out our main argument, we will also study the case where the invariance assumption does not hold.
\begin{align}\label{thm:power-3}
P( \Xin > \tau/2 )  & \le \sum_{r\neq r'} P\big( | t_n(G_r\eps) - t_n(G_{r'} \eps) |  > \tau/2\big)  +  \sum_{r} P\big( | t_n(G_r\eps) - t_n( \eps) |  > \tau/2\big) 
\nn\\
& \myle{(i)} \sum_{r \neq r'} P\big( |\Lambda_n|  > \tau/2)
+  \sum_{r}P\big( |\Lambda_n| > \tau/2\big)
  \nn\\
& = O(m^2) P\big( |\Lambda_n|  > \tau/2 \big) .
\end{align}
Here, (i) follows from i.i.d. $G_r$ and the invariance of $\eps$. 
Note that 
$\sigma^2_{\Lambda_n} = \mE(\Lambda_n^2) =  \mE[(t_n(G_1 \eps) - t_n(G_2\eps))^2]$ and so,
\begin{equation}\label{808}
\sigma^2_{\Lambda_n} =  2\mE[t_n^2(G \eps)] - 2 [\mE(t_n(G_1\eps) t_n(G_2\eps))] = 2\var(t_n(\eps)) - 2\var(\nu(\eps)),
\end{equation}
where $\nu(\eps) = \mE[ t_n(\eps)|\eps ]$.
In other terms, $\sigma_{\Lambda_n^2} = 2\mE[\mVar(t_n(G\eps) \mid \eps) ]$ --- this is denoted as $\sigma_n^2$ in the theorem.
It follows that
\begin{align}\label{thm:power-3}
P( \Xin > \tau/2 )  = O(m^2) P\big( |\Lambda_n|  > \tau/2\big)  
\le O(m^2) \frac{4\mE(\Lambda_n^2)}{\tau^2}
= O(m^2) \frac{\sigma^2_{\Lambda_n}}{\tau^2},
\end{align}
from Chebyshev's inequality for any $\tau > 0$.
 Now, let $\Tne^\tau = \{t_n(\eps) + \tau\} \cup \{t_n(G_r\eps) \}$ denote the 
randomized values under the alternative.
We claim that
\begin{equation}\label{eq3028}
 c_{n,\alpha}(\Tne^\tau) \le c_{n,\alpha^-}(\Tne)
\end{equation}
 for a specific value of $\alpha^- \in [0, 1)$. 
  That is, 
the critical value under $\Tne^\tau$ is still some critical value within $\Tne$. 
Intuitively, this is true since the critical value of some $S$ remains robust to a ``contamination" of just one single value in $S$.
Formally, since the critical value on set $S$
 is an ordered statistic of $S$, it follows that 
 $c_{n,\alpha}(\Tne^\tau) \neq c_{n,\alpha}(\Tne)$ only when 
 $$t_n(\eps) \le c_{n,\alpha}(\Tne),~\text{and}~t_n(\eps) + \tau > c_{n,\alpha}(\Tne).
 $$
 This implies that $c_{n,\alpha}(\Tne^\tau)$ is at most equal to the ordered statistic that is just larger than $c_{n,\alpha}(\Tne)$ within $\Tne$. Let $R_n$ be the multiplicity of test statistic values on this ordered statistic. Then, the new critical value satisfies
\begin{equation}\label{eq3056}
c_{n,\alpha}(\Tne^\tau) \le c_{n,\alpha}(\Tne) + R_n/(m+1) 
= c_{n, \alpha^-}(\Tne),
\end{equation}
by setting $\alpha^- = \alpha - R_n /(m+1)$, thus satisfying~\eqref{eq3028}. 
 In fact, the analysis in Lemma~\ref{lemma:xi} shows that $R_n = 1 + o_P(1)$ 
 and so $a^- = \alpha - O(1/m) +  o_P(1/m) \in(0, 1)$ for large enough $m$.

Under the alternative hypothesis,
$T_n = t_n(\epsn) + \tau$ for some $\tau > 0$, and so
\begin{align}\label{thm:power-4}
\Ind\{T_n > c_{n,\alpha}(\Tner ) \} & \myge{(i)} 
\Ind\{T_n > c_{n,\alpha}(\Tne^\tau) + \DeltaMax\} \nn\\
& \myge{(ii)} \Ind\{T_n > c_{n,\alpha^-}(\Tne) + \DeltaMax\} \nn\\
& \myeq{(iii)} \Ind\{ t_n(\eps)  >    c_{n,\alpha^-}(\Tne) + \DeltaMax  -\tau\}
\nn\\
& = \Ind\{ t_n(\eps)  >    c_{n,\alpha^-}(\Tne)  - \Xin + \DeltaMax  -\tau + \Xin\}
\nn\\
& \myge{(iv)} \Ind\{ t_n(\eps)  \ge  c_{n,\alpha^-}(\Tne) -\Xin \} \cdot \Ind\{ \DeltaMax  +\Xin  < \tau \}
\nn\\
& \myeq{(v)} \Ind\{ \DeltaMax  +\Xin < \tau \}
\nn\\
& \myge{(vi)} \Ind\{ 2\Xin \le \tau \} - \Ind\{\Delta_n \ge \Xin\}
\nn\\
  P\{T_n > c_{n, \alpha}(\Tner) \}
& \ge P(\Xin \le  \tau/2) - P(\Delta_n \ge \Xin)
\nn\\
& \myge{(vii)} 1 - O(m^2) \frac{ \sigma^2_{\Lambda_n}  }{\tau^2} - O(m^2) (\gon + 1/|\Gall|) - A(\gamma, m) \cdot O(\zeta_n^{2\gamma/(2+\gamma)}).
\end{align}
In the above derivations, result (i) follows from 
the same argument as in derivation (i) of~\eqref{eq:one-sided_bound};
(ii) follows from~\eqref{eq3028};
(iii) follows from the alternative hypothesis; 
(iv) follows from the indicator identity $\Ind\{a>  b\} \ge \Ind\{a \ge 0\}\Ind\{b < 0\}$, by setting 
$a = t_n(\eps) - c_{n,\alpha^-}(\Tne) + \Xin$ 
and $b = \DeltaMax + \Xin - \tau$;
result (v) follows from~\eqref{thm:power-1}; 
(vi) follows from indicator identity $\Ind\{d+\xi < \tau\} + \Ind\{d \ge \xi\}  \ge \Ind\{2\xi \le \tau\}$; 
finally, (vii) follows from~\eqref{thm:power-3} combined with~\eqref{thm:power-2}.

We see that the test has power one asymptotically for a sequence of alternatives $\sigma_{\Lambda_n} / \tau \to 0$ and $\tau>0$.
Under the cdf lower bound in the theorem,~\eqref{thm:power-3} can be further analyzed as follows:
\begin{align}\label{thm:power-333}
P( \Xin > \tau/2 )   & = O(m^2) P\big( |\Lambda_n|  > \tau/2 \big) = 
 O(m^2) P\big( |\Lambda_n| /\sigma_{\Lambda_n}  > \tau/2 \sigma_{\Lambda_n} \big) \nn\\
& =  O(m^2) \big\{1 - [ \FLn(\tau/2\sigma_{\Lambda_n}) - \FLn(-\tau /2\sigma_{\Lambda_n}) ] \big\} \nn\\
& = O(m^2) e^{-q(\tau/2\sigma_{\Lambda_n}) } =  O(m^2) e^{-q(\tau/2\sqrt{2}\sigma_{\Lambda_n}) }.
\end{align}
We can now plug-in this bound in the last step of~\eqref{thm:power-4} to get a lower bound for the power.

For the case where the invariant hypothesis does not hold, note that the derivation 
of Equation~\eqref{thm:power-4} still implies that the test has asymptotic power 1 provided that the term $P\big( | t_n(G_r\eps) - t_n( \eps) |  > \tau/2\big)$ that appears in the derivation of Equation~\eqref{thm:power-3} converges to 0 as  $\tau\to\infty$.

\end{proof}

\section{Proofs of Theorems~4-5}
\subsection*{Notation and Preliminaries}\label{appendix:notation2}
Our working model is the linear regression model:
\begin{equation}\label{eq:linear_model}
y = X\beta + \eps.
\end{equation}
As $X$ are non-stochastic, all probability statements are conditional on $X$. 

For a set $S\subseteq \{1, \ldots, n\}$ 
let $\ones_S\in\Real^n$ denote the  binary vector that is equal to one at $i$-th element only if $i\in S$, and is zero otherwise.
We will use $\ones_i$  as a shorthand for $\ones_{\{i\}}$. $\U = \ones \ones'$ will denote the $n\times n$ matrix of ones, 
and $\Id = \sum_{i=1}^n \ones_i \ones_i'$ is the $n\times n$ identity matrix. We will use subscripts on these matrices as necessary when the dimension is different than $n$. We also use $\lambda_i(A)$ to denote the $i$-th eigenvalue of t a positive-definite matrix $A$, 
and $\lmax(A)$ and $\lmin(A)$ for the maximum and minimum eigenvalues, respectively. The condition number of $A$ is
 $\kappa_n(A) = \lmax(A)/\lmin(A)$. We will use boldface $\mathbf{A}$ mainly to 
 emphasize that a matrix is positive-definite ---usually a variance-covariance matrix---, i.e., $\lmin(\mathbf{A}) > 0$.

For a positive definite matrix $A\in\Real^{n\times n}$, the following result will be useful:
\begin{equation}\label{eq:prelim1}
|| A ||^2_F \le  n \lmax^2(A).
\end{equation}
For a quick proof, note that $\lmax^2(A)$ is the maximum eigenvalue of $A^\top A$.
Thus, $||A||^2_F = \trace(A^\top A) = \sum_{i=1}^p \ones_i' (A^\top A) \ones_i \le \lmax^2(A) \sum_i ||\ones_i||^2
=  n \lmax^2(A)$. Moreover, 
\begin{equation}\label{eq:prelim2}
\trace(A) =  \sum_{i=i}^n \lambda_i(A)  \le  n\lmax(A).
\end{equation}
A useful corollary is
\begin{equation}\label{eq:prelim3}
\trace(\XnT X) \trace\{(\XnT X)^{-1}\} \le p^2 \lmax(\XnT X) / \lmin(\XnT X) = p^2 \kappa_n,
\end{equation}
where $\kappa_n$ denotes the condition number of $\XnT X$.
We will use $\xj\in\Real^n$ to denote the $j$-th column of $X$, 
and $\xrow{i}$ to denote the $i$-th row of $X$. 
We will sometimes write $X$ in column form as $X = [x_{\cdot 1}; \ldots; x_{\cdot p}]$.
For the linear model, we let
\begin{equation}\label{def:Q}
\Qx = (\XnT X)^{-1} X^\top,
\end{equation}
where $\Qx$ is the $p\times p$ matrix such that $\hbeta = \Qx y = \beta + \Qx \eps$,  and $t_n(\eps) = a' \Qx \eps$. Vector $a$ is equal to $e_j$ when we test the individual significance hypothesis, $H_{0,j}$.

The following property of $\Qx$ will be useful:
\begin{align}\label{eq:Q_prop}
\Qx \Qx^\top  & = (X^\top X)^{-1}.
\end{align}
Moreover, since the first column of $X$ is ones, it holds $X \ones_1 = \ones$ and so
$ (\XnT X)^{-1} \XnT \ones = \ones_1$. This implies
\begin{align}\label{eq:Q_prop2}
\Qx \U \Qx^\top = \ones_1 \ones_1'.
\end{align}
We also let $\Px = X (\XnT X)^{-1} \XnT = X \Qx$ denote the ``hat matrix" for which $\hat y = X(\hbeta-\beta) = X \Qx \eps = \Px \eps$. This matrix has the following useful properties:
\begin{align}\label{eq:Px_prop}
\Px^\top & = \Px,~\text{and}~\Px \Px^\top  = \Px,\nn\\
 \trace(\Px) & = \trace\{X (\XnT X)^{-1} \XnT\} =  \trace\{ (\XnT X)^{-1} \XnT X\} = \trace(\Id_{p\times p}) = p,\nn\\
 \Px \U & = X \Qx \ones\ones'= X \ones_1\ones' = \ones \ones' = \U,\nn\\
 ||\Px z||^2 & = z' (\Px^\top \Px) z = z' \Px z \le \lmax(\Px) ||z||^2 \le \trace(\Px)  ||z||^2 = p||z||^2,~~z\in\Real^n.
\end{align}
For the restricted OLS estimator we will use the matrix
\begin{equation}\label{eq:def_Px0}
\Pxo = X (\Id - \S a a'/a' \S a) \Qx = \Px - X \S aa' \S \XnT/ (a'\S a),
\end{equation}
where $\S = (\XnT X)^{-1}$. Note that $\S$ is positive definite by Assumption~\AIb. This is still a projection matrix since the following properties hold:
\begin{align}\label{eq:Pxo_prop}
(\Pxo)^\top & = \Pxo,~\text{and}~\Pxo (\Pxo)^\top  = \Pxo,\nn\\
 \trace(\Pxo) & =\trace(\Px) - \trace(X \S a a' \S \XnT/a' \S a) = p - \trace(a' \S \XnT X \S a) / a' \S a = p-1,\nn\\
 \Pxo \U & = \Px \U - X \S a a' \S \XnT \ones \ones' / (a' \S a) = \U - a_1 X \S a \ones' / a'\S a.
\end{align}

\newcommand{\bTr}{\mathbf{T}_{\mathrm r}}
\newcommand{\bTTr}{\mathbf{\tilde T}_{\mathrm r}}

\begin{lemma}\label{lemma:C_bounds_tight}
For $G\in\Unif(\Gall)$, define $\V(z) = \mE(G z z' G^\top)$, $z\in\Real^n$. 
Suppose that $\V(z)$ can be written as
 $$
\V(z) = \sum_{k=1}^K h_k(z) \A_k,
$$ 
where 
\begin{enumerate}[(a)]
\item $\A_k\in\Real^{n\times n}$ are constant (parameter-free) positive-definite matrices, and
\item  $h_k: \Real^n\to\Real$ with 
$h_k(z) = \sum_{j=1}^K c_{kj} \cdot (z' \A_j z)$ for some constants $c_{kj}\in\Real$.
\end{enumerate}
Let $\mu_k  = \mE\big(h_k(\eps)\big)$, $\tilde\mu_k = \mE\big(h_k(\Px\eps)\big)$ and $\psi_k  = a' \Qx \A_k \Qx^\top a \ge 0$.
Also, let $\nu(\eps) = \mE(t_n(G\eps) | \eps)$, and suppose that there exist $\rho_k\in\Real$ such that
\begin{equation}\label{eq:def_nu}
\mE[\nu^2(\eps)] = \sum_{k=1}^K \rho_k \mu_k \psi_k.
\end{equation}
Then, in the linear model and the randomization test with $t_n(\eps) = a'(\XnT X)^{-1} \XnT\eps$ as the statistic, Condition (C1) is reduced to
\begin{equation}\label{eq:psi_condition}
\frac{\sum_{k=1}^K \tilde\mu_k \psi_k}{\sum_{k=1}^K(1-\rho_k) \mu_k \psi_k}  \to 0.
\end{equation}
Moreover, let us define the $K\times K$ matrices 
$\bTr = [\trace(\A_k \A_j)]$, $C= [c_{kj}]$ and $\bTTr = [\trace(\Px^\top \A_k \Px \A_j)]$, for $k,j=1, \ldots, K$. 
If $\eps$ satisfy the invariance property for all $n>0$, such that $\eps\myeq{d}\gg\eps$ for all $\gg\in\Gall$, 
and $\mE(\eps|X) = 0$, then 
$$
C = \bTr^{-1},~\text{and}~\tilde \mu = C \bTTr \mu.
$$
For the restricted OLS residuals, we just replace $\Pxo$ instead of $\Px$ in the definition of $\bTTr$ and $\tilde\mu_k$.
\end{lemma}
\begin{proof}
In the linear model, $t_n(\eps) = a' \Qx \eps$, and so using regular residuals we obtain
\begin{align}\label{lemma-bounds-tight-3}
\Delta_n^2 &= (t_n(G\heps) - t_n(G\eps))^2 = [a' \Qx G(\heps - \eps)]^2 \nn\\
\Delta_n^2  & \myeq{(i)} a' \Qx G \Px \eps \eps' \Px^\top G^\top \Qx^\top a \nn\\
\mE(\Delta_n^2) & = a' \Qx \mE(G \Px \eps \eps' \Px^\top G^\top) \Qx^\top a \myeq{(ii)} a' \Qx \mE\big(\V(\Px \eps)\big) \Qx^\top a \nn\\
& = \sum_{k=1}^K  \tilde\mu_k (a' \Qx \A_k \Qx^\top a) = \sum_k \tilde\mu_k \psi_k.
\end{align}
Here, (i) uses $\heps - \eps =- \Px\eps$ from regular OLS; (ii) follows by definition of $\V(z)$, and (ii) follows from~\eqref{lemma-bounds-tight-2}.

From the definition of $\Lambda_n$,  we also obtain
\begin{align}\label{lemma-bounds-tights-4}
\mE(\Lambda_n^2) &= 2 \mE\{[t_n(G\eps)]^2\}  - 2 \mE[t_n(G_1\eps) t_n(G_2\eps)] 
= 2a' \Qx \mE[\V(\eps)] \Qx^\top a - 2 \mE[\nu^2(\eps)]\nn\\
& = 2\sum_k \mu_k \psi_k -  2 \sum_k \rho_k \mu_k \psi_k 
= 2\sum_k (1-\rho_k)  \mu_k\psi_k.
\end{align}
Condition~\cVar~is now reduced to
$$
\zeta_n^2 = \frac{\mE(\Delta_n^2)}{\mE(\Lambda_n^2)} =
\frac{ \sum_k \tilde\mu_k \psi_k}{2 \sum_k (1-\rho_k) \mu_k \psi_k}  \to 0.
$$

For the second part of the theorem, note that under the invariance hypothesis, it holds that $\mE[\V(\eps)] = \sum_k \mu_k \A_k$ but also
$$
\mu_k = \mE\big(h_k(\eps)\big) = \sum_{j=1}^K c_{kj} \mE(\eps' \A_j \eps) = \sum_j c_{kj}  \trace(\A_j \Ve) =  
\sum_j c_{kj} \sum_{\ell=1}^K \mu_\ell \trace(\A_j \A_\ell).
$$
From this, we obtain $\mu = C \bTr \mu$, where $\mu = (\mu_1, \ldots, \mu_K)$, and so $C=  \bTr^{-1}$. Similarly, we obtain
\begin{equation}\label{lemma-bounds-tight-2}
\mE\big(\V(\Px\eps)\big) = \sum_{k=1}^K \mE(h_k(\Px\eps)) \A_k =  \sum_k \tilde\mu_k \A_k,
\end{equation}
and 
$$
\tilde \mu_k = \mE\big(h_k(\Px \eps)\big) = \sum_{j=1}^K c_{kj} \mE(\eps' \Px^\top \A_j \Px \eps) = 
 \sum_j c_{kj}  \sum_{\ell} \mu_\ell \trace(\Px^\top \A_j \Px \A_\ell),
$$
from which we obtain $\tilde\mu =C  \bTTr  \mu$. For restricted OLS residuals,
the expressions are analogous and use $\Pxo$ instead of $\Px$.

\end{proof}

\subsection{Proof of Theorem 4}
\begin{proof}
In this setting, $G\sim\Unif(\Gall^\pp)$ can be represented as $G =\sum_{i=1}^n \ones_{\pi(i)} \ones_i'$ where $\pi$ is a random permutation.
Then,
\begin{align}\label{eq:perm-ones}
\mE(\ones_{\pi(i)})& = (1/n) \ones,~\text{and so}~\mE(G) = (1/n) \ones \sum_i \ones_i' = \U/n;\nn\\
\mE(\ones_{\pi(i)} \ones_{\pi(i)}' ) & = (1/n) \sum_{i=1}^n \ones_i \ones_i' = (1/n) \Id, \nn\\
\mE\big (\ones_{\pi(i)} \ones_{\pi(j)}' \mid i\neq j\big)  &= 
\mE\big [\ones_{\pi(i)} \mE(\ones_{\pi(j)}' \mid \pi(i), i\neq j) \mid i\neq j\big] \nn\\
& = \mE\big [\ones_{\pi(i)} (\ones-\ones_{\pi(i)})' /(n-1) \mid i\neq j\big] =
(\U - \Id) / n(n-1).
\end{align}

For the covariance matrix function, 
\begin{align}\label{eq:perm-Vz}
\V(z) = \mE(G z z' G^\top) & =
  \mE\bigg\{ (\sum_{i=1}^n \ones_{\pi(i)} \ones_i') z z'  (\sum_{j=1}^n \ones_j \ones_{\pi(j)}')      \bigg\} \nn\\
& = \sum_{i, j} z_i z_j\mE(\ones_{\pi(i)} \ones_{\pi(j)}') = 
\sum_{i=1}^n z_i^2 \mE(\ones_{\pi(i)} \ones_{\pi(i)}') +  \sum_{i\neq j} z_i z_j \mE(\ones_{\pi(i)} \ones_{\pi(j)}')  \nn\\
& \myeq{(ii)} (||z||^2/n) \Id + [n^2 (\bar z)^2 - ||z||^2] (\U - \Id) /n(n-1) \nn\\
& = s^2(z) (\Id - \U/n) + (\bar z)^2 \U,
\end{align}
where $s^2(z) = [ ||z||^2 - n(\bar z)^2] / (n-1)$ is the sample variance in vector $z$. 

Next, we derive all the quantities that are necessary for the application of Lemma~\ref{lemma:C_bounds_tight}. 
First, note that $\V(z)$ can be represented through the following basis~($K=2$): 
\begin{align}\label{eq:perm-A}
\A_1 & = (\Id - \U/n),~\A_2 = \U,\nn\\
h_1(z) & = s^2(z) = [1/(n-1)] z' \A_1 z,~h_2(z) = (\bar z)^2 = (1/n^2) z' \A_2 z.
\end{align}

Therefore, 
$$
\psi_1 = a' \Qx \Qx^\top a = a' (\XnT X)^{-1} a,~\text{and}~
\psi_2 = a' \Qx \U \Qx^\top a = 0,
$$
since the columns of $X$ are centered. Moreover, 
$$
\mu_1 = \mE[s^2(\eps)] = \mE(||\eps||^2)/ (n-1),~\text{and}~
\mu_2 = \mE[(\bar\eps)^2] = 0,
$$
since $\eps_i$ are centered as well. Next, we obtain
$$
\tilde\mu_1 = \mE[s^2(\Px\eps)] = \mE(||\Px\eps||^2)/ (n-1).
$$
Finally, we see that $\rho_1, \rho_2=0$ since $\mE[t_n(G\eps)|\eps] = a'\Qx \XnT (\bar\eps \ones) = 0$. Applying Lemma~\ref{lemma:C_bounds_tight}, we can conclude that Condition~(C1) reduces to
\begin{equation}\label{eq:C1_perm}
\frac{ \mE(||\Px \eps||^2)}{\mE(||\eps||^2)} \to 0.
\end{equation}
We now consider two cases.

\vspace{5px}
\noindent\underline{Invariant hypothesis is true.}
Here, we assume that the invariant hypothesis is true under $\Gall^\pp$, such that 
$
\eps \myeq{d} \gg \eps | X,~\text{for all permutations}~\gg\in\Gall^\pp.
$
Then, for $G\sim\Unif(\Gall^\pp)$,
$$
\mE(\eps\eps'|X) = \mE(G\eps \eps G^\top | X) = \mE(||\eps||^2)/(n-1) \Id.
$$
This leads to $\mE(||\Px \eps||^2) =  \mE(\eps' \Px \eps) = \trace\{\Px \mE(\eps \eps')\} = \mE(||\eps||^2) p/(n-1)$. Thus, under the invariant hypothesis, the limit in Equation~\eqref{eq:C1_perm} becomes
\begin{equation}\label{eq1159}
    \frac{ \mE(||\Px \eps||^2)}{\mE(||\eps||^2)} = p/n + o(1) \to 0.
\end{equation}

\vspace{5px}
\noindent\underline{Limit hypothesis is true.}
Under this assumption, we can now leverage the results of~\citet{diciccio2017robust} who showed that, for $G\sim\Unif(\Gall^\pp)$,  the limit hypothesis implies
$$
\frac{t_n(G\eps)}{\sigma_n(G\eps)} \limeq{d} N(0,1).
$$
Moreover, they showed that $\sigma_n^2(G\heps)/\sigma_n^2(G\eps) \limeq{p} 1$, 
while $n\sigma_n^2(G\eps) \limeq{p} \sigma^2 > 0$.

To finish the proof, we turn our focus to working out Condition (C1) for the studentized statistic, $t_n^s(\eps) = t_n(\eps)/\sigma_n(\heps)$. The condition can be calculated as

\begin{equation}\label{eq1183}
\frac{\mE[(t_n^s(G\heps) - t_n^s(G\eps))^2]}{\mE[(t_n^s(G\eps))^2]}
= \frac{\mE[ (1/n\sigma_n^2(G \heps)) (t_n(G\heps) - t_n(G\eps))^2]}{
\mE[ (1/n\sigma_n^2(G \heps)) t_n^2(G\eps) ] }.
\end{equation}
Since $n\sigma_n^2(G\heps)$ converges in probability, the ratio in~\eqref{eq1183} converges to 0 if Condition (C1) holds true for $t_n(\eps)$, i.e., the limit condition in Eq.~\eqref{eq:C1_perm} is true.
However, the limit condition in Eq.~\eqref{eq:C1_perm} indeed converges to 0 under the standard OLS assumptions of the limit hypothesis. This concludes the proof.
\end{proof}

\subsection{Proof of Theorem 5}
\begin{proof}
In this setting,  we can represent $\GG\sim\Unif(\Gall^\ss)$ as $\GG = \sum_{i=1}^n  S_i \ones_i \ones_i'$, 
where $S_1, \ldots, S_n$ are independent random signs. 
For any vector $z\in\Real^n$, we thus get
\begin{align}\label{eq:s-Vz}
\V(z) = \mE(G z z' G^\top) & =
\mE\big\{  (\sum_{i=1}^n S_i \ones_i\ones_i') z z' (\sum_{j=1}^n S_j \ones_j\ones_j')  \big\}
 =  \mE\big(\sum_{i, j}  S_i S_j z_i z_j \ones_i \ones_j'\big)  \nn\\
& \myeq{(i)} \sum_{i} \mE(S_i^2) z_i^2 \ones_i \ones_i' \myeq{(ii)} \sum_i z_i^2 \ones_i \ones_i' =\diag_n(z_i^2).
\end{align}
Here, (i) follows from independence of $S_i, S_j$ for $i\neq j$; and (ii) follows from $E(S_i^2)=1$. 
Hence, $\V(z)$ is a diagonal matrix 
with $z_i^2$ being the $i$-th diagonal element.

We start with regular OLS residuals. In the notation of Lemma~\ref{lemma:C_bounds_tight}, $\V(z)$ can be written using the following basis~($K=n$): 
\begin{align}\label{eq:s-A}
\A_k = \ones_k\ones_k',~\text{and}~h_k(z) = z_k^2=z'\A_k z,~k=1, \ldots, n.
\end{align}
We thus obtain
$$
\mu_k =\mE[h_k(\eps)]= \mE(\eps_k^2),
$$
and 
\begin{align}\label{eq:s-psi}
\psi_k & = a' \Qx \ones_k\ones_k' \Qx^\top a  = [\Qx^\top a a' \Qx]_{kk}
\end{align}
Moreover,
$$
\tilde\mu_k = \mE[h_k(\Px\eps)]= \sum_{i=1}^n h_{ki}^2 \mE(\eps_i^2) =  \sum_{i=1}^n h_{ki}^2 \mu_i.
$$
Note that $\mu_k, \tilde\mu_k, \psi_k \ge 0$ and $\sum_k \psi_k = a' \Qx \Qx^\top a = a' \S a >0$.
Finally, for the conditional mean we obtain 
$\nu(\eps) = \mE[t_n(G\eps) | \eps] = a' \Qx \mE(G) \eps = 0$, which implies that $\rho_k=0$ 
Now we are ready to apply the lemma.

Let $\mu^* = \max_k \mu_k$ and  $\mu_* = \min_{k} \mu_k$. Then,
$$
\tilde\mu_k = \sum_i \mu_i  h_{ki}^2 \le \mu^*   \sum_i h_{ki}^2
= \mu^*  || \ones_k' \Px ||^2 \myeq{(i)}  \mu^* \ones_k' \Px \ones_k
\le \mu^*  h_{kk} \le \mu^*  \lev p/n,
$$
where (i) follows from~\eqref{eq:Px_prop}. Then, the ratio in Lemma~\ref{lemma:C_bounds_tight} is equal to
\begin{align}\label{eq:s-ratio}
\frac{ \sum_k \tilde\mu_k \psi_k}{2 \sum_k (1-\rho_k) \mu_k \psi_k}
 \le \frac{   (\mu^* \lev p/n) \sum_k \psi_k}{ 2 \mu_* \sum_k \psi_k}
 = \frac{\mu^*}{\mu_*}  \lev p/n.
\end{align}
We can also obtain a similar bound by considering $\psi^* = \max \psi_k$ and $\psi_* = \min\psi_k$. 
Since all $\mu_k \ge 0, \psi_k \ge 0$, we have 
$$
\sum_k \tilde\mu_k \psi_k \le \psi^* \sum_k\tilde \mu_k =  \psi^* \sum_k \sum_i \mu_i h^2_{ki} 
=  \psi^* \sum_i \mu_i  \sum_k  h^2_{ki} \le  \psi^* (\lev p /n) \sum_i \mu_i .
$$
Moreover,
$$
\sum_k (1-\rho_k) \mu_k \psi_k = \sum_k \mu_k \psi_k \ge \psi_* \sum_k \mu_k.
$$
Combining these two results, we obtain another bound for~\eqref{eq:s-ratio}:
\begin{align}\label{eq:s-ratio-2}
\frac{ \sum_k \tilde\mu_k \psi_k}{2 \sum_k (1-\rho_k) \mu_k \psi_k}
 \le \frac{\psi^*}{\psi_*}  \lev p/n.
\end{align}

Thus, a sufficient condition for asymptotic validity is
\begin{equation}\label{eq:s-final}
\min\{ \mu^*/\mu_*, \psi^*/\psi_*\}  ~\lev ~p/n = o(1).
\end{equation}

Regarding the restricted OLS estimator,~\eqref{eq:def_Px0} implies
\begin{align}\label{eq:s-TTr0}
(\Pxo)_{kk} = h_{kk} - (a' \S \XnT \ones_k)^2/a'\S a \le h_{kk} \le \lev p/n.
\end{align}
This means that the same bound as in~\eqref{eq:s-ratio} holds for restricted OLS as well.

\vspace{5px}
\noindent\underline{Invariant hypothesis is true.}
Note that in contrast to Theorem 4, where the main condition simplified under the invariant hypothesis, 
the condition under sign symmetry does not lead to a similar simplification. 

\vspace{5px}
\noindent\underline{Limit hypothesis is true.}
We can follow an identical approach as in the proof of Theorem 4. 
The necessary asymptotic results on the randomization distribution of $t_n^s(G\eps)$, where $G$ are random sign flips, have been established by ~\citet{janssen1999nonparametric}. 
Following the proof of Theorem 4 we can use these results to show that if Condition~\eqref{eq:s-final}  holds true for $t_n(\eps)$, then it should also hold true for its studentized version $t_n^s(\eps)$.

\end{proof}

\end{document}